\documentclass[fleqn,usenatbib]{mnras}

\usepackage[T1]{fontenc}
\usepackage{ae,aecompl}

\usepackage{graphicx}	% Including figure files
\usepackage{amsmath}	% Advanced maths commands
\usepackage{amssymb}	% Extra maths symbols
\usepackage{hyperref}
\usepackage{chngcntr}
\counterwithout{figure}{section}
\counterwithout{table}{section}
\usepackage{nccmath}
\usepackage{placeins}
\usepackage{adjustbox}
\usepackage{array}
\usepackage{ulem}
\usepackage{xcolor}
\title[Cosmology and baryons on small scales]{Exploring extensions to the standard cosmological model and the impact of baryons on small scales}
\author[S. Stafford et al.]{
Sam G. Stafford$^{1}$\thanks{E-mail: s.stafford@2014.ljmu.ac.uk}, Shaun T. Brown$^1$, Ian G. McCarthy$^{1}$\thanks{E-mail: i.g.mccarthy@ljmu.ac.uk}, Andreea S. Font$^1$, \newauthor Andrew Robertson$^2$, Robert Poole-McKenzie$^1$
\\
$^{1}$Astrophysics Research Institute, Liverpool John Moores University, 146 Brownlow Hill, Liverpool L3 5RF, UK \\
$^{2}$Institute for Computational Cosmology, Durham University, South Road, Durham DH1 3LE, UK\\
}

\date{Accepted 2020 July 7. Received 2020 July 1; in original form 2020 April 8.}

\pubyear{2020}

\begin{document}
\label{firstpage}
\pagerange{\pageref{firstpage}--\pageref{lastpage}}
\maketitle

\begin{abstract}
It has been claimed that the standard model of cosmology ($\Lambda$CDM) cannot easily account for a number of observations on relatively small scales, motivating extensions to the standard model.  Here we introduce a new suite of cosmological simulations that systematically explores three plausible extensions: warm dark matter, self-interacting dark matter, and a running of the scalar spectral index of density fluctuations.  Current observational constraints are used to specify the additional parameters that come with these extensions.  We examine a large range of observable metrics on small scales, including the halo mass function, density and circular velocity profiles, the abundance of satellite subhaloes, and halo concentrations.  For any given metric, significant degeneracies can be present between the extensions.  In detail, however, the different extensions have quantitatively distinct mass and radial dependencies, suggesting that a multi-probe approach over a range of scales can be used to break the degeneracies.  We also demonstrate that the relative effects on the radial density profiles in the different extensions (compared to the standard model) are converged down to significantly smaller radii than are the absolute profiles.  We compare the derived cosmological trends with the impact of baryonic physics using the \texttt{EAGLE} and \texttt{ARTEMIS} simulations.  Significant degeneracies are also present between baryonic physics and cosmological variations (with both having similar magnitude effects on some observables).  Given the inherent uncertainties both in the modelling of galaxy formation physics and extensions to $\Lambda$CDM, a systematic and simultaneous exploration of both is strongly warranted.
\end{abstract}


%%%%%%%%%%%%%%%%%%%%%%%%%%%%%%%%%%%%%%%%%%%%%%%%%%

%%%%%%%%%%%%%%%%% BODY OF PAPER %%%%%%%%%%%%%%%%%%

\section{Introduction}
The current cosmological paradigm describes a universe which has a matter content that is primarily composed of collisionless cold dark matter (CDM), a cosmological constant ($\Lambda$), and normal baryonic matter \citep{PlanckXIII}. 
The true nature of dark matter, however, is still an unsolved problem.  Aside from there not yet being a confirmed direct detection of dark matter, there are also claimed problems with the current cosmological model (most commonly abbreviated to the $\Lambda$CDM model) on small scales which could hint at a deviation from CDM.  For example, three of the most widely discussed issues with this model are the `cusp--core problem' \citep{Flores1994, Moore1994}, the `missing satellites problem' \citep{Klypin1999, Moore1999}, and the `too-big-to-fail problem' \citep{Boylan-Kolchin2011}. A recent review on these, along with other apparent small-scale issues identified within $\Lambda$CDM, can be found in \cite{Bullock2017}.

These issues have fuelled research into alternatives/extensions to the standard model, particularly modifications to the nature of dark matter.  For example, one can relax the assumption that dark matter is purely collisionless, allowing instead for a non-negligible self-interaction with a cross-section that can be constrained by observations.  Previous studies on such self-interacting dark matter (SIDM) suggest it may be able to solve the cusp--core problem \citep{Spergel2000, Yoshida2000, Dave2001, Colin2002,  Vogelsberger2012, Rocha2013, Elbert2015, Kaplinghat2016}, the too-big-to-fail problem (provided that the self-interaction cross-section is large enough e.g. \citealt{Zavala2013, Elbert2015}), and may help to explain the observed diversity of galaxy rotation curves \citep{Kamada2017, Creasey2017}. 

Another popular extension retains the assumption that dark matter is collisionless, but invokes a non-negligible thermal velocity at early times.  Termed warm dark matter (WDM), the associated free-streaming erases small-scale density perturbations leading to a characteristic cut-off in the linear matter power spectrum \citep{Bond1980, Pagels1982, Dodelson1994, Hogan2000, Viel2005, Abazajian2006}.  This leads to WDM potentially being able to resolve the missing satellites problem due to the suppression of structure with masses close to the cut-off scale in the matter power spectrum \citep{Colin2000, Bode2001, Polisensky2011, Lovell2012, Anderhalden2013, Bozek2016, Horiuchi2016, Bose2017}. Furthermore, due to the later formation times of haloes in a WDM cosmology, haloes tend to have a lower central density which helps to mitigate the too-big-to-fail problem \citep{Lovell2012, Horiuchi2016, Lovell2017a}. Studies have also shown that WDM can produce cores in the density profile of dark matter haloes, however, these are not large enough to solve the cusp--core problem given current constraints on the mass of the dark matter particle \citep{Villaescusa-Navarro2011, Macci2012, Shao2013}.

An alternative mechanism that can alter small-scale structure but which treats dark matter in the standard way (cold and collisionless), is to invoke a change in the primordial power spectrum on small scales.  The standard model of cosmology assumes that the primordial power spectrum of density fluctuations, $P(k)$, laid down by inflation is a pure power-law.  Measurements of the cosmic microwave background constrain the power-law exponent to be $n_s \approx 0.96$, in excellent agreement with generic inflation models.  However, even the simplest single-field inflation models predict some degree of deviation from a pure power-law (\citealt{Kosowsky1995}, see also \citealt{Garcia-Bellido2014, Escudero2016}). 
 
A more general treatment of the primordial matter power spectrum allows the spectral index to vary with scale, with this scale-dependence termed the `running' of the spectral index ($\alpha_s$). 
As discussed in \citet{Stafford2020}, a growing body of observational evidence suggests some preference for a mildly negative value for $\alpha_s$ \citep{Dunkley2011, Zhao2013, Hou2014, PlanckXIII, Palanque-Delabrouille2015, palanque-delabrouille2019}, which will act to suppress power on the smallest (and largest) of scales.  \citet{Garrison-Kimmel2014} have previously shown that a negative running can help resolve the missing satellite problem as well as alleviate the too-big-to-fail problem.

It is possible, however, that the small-scale crises outlined previously which supposedly exist with the $\Lambda$CDM model are instead just a product of conclusions drawn from comparing the observable Universe to the predictions from a series of dark matter-only simulations.   Recently there has been extensive research into the possibility that a realistic treatment of baryonic physics in simulations, such as supernovae feedback, stellar winds, and feedback from active galactic nuclei (AGN), may be able to resolve some of these tensions.
These studies have shown the importance of including baryonic physics in the simulations, demonstrating the potential to resolve for the cusp--core problem \citep{Mashchenko2008, Pontzen2012, Maccio2012, Madau2014, Onorbe2015, Read2016, Tollet2016, Wetzel2016, Fitts2017}, the missing satellites problem \citep{Bullock2000, Benson2002, Somerville2002, Kravtsov2004a, DOnghia2010, Wetzel2016, Garrison-Kimmel2017, Sawala2017}, as well as the too-big-to-fail problem \citep{Zolotov2012, Arraki2014, Brooks2014, Chan2015, Wetzel2016, Tomozeiu2016, Sawala2016, Dutton2016, Brooks2017} and without having to invoke any extensions to the standard model.  

To what extent the successes of hydrodynamical simulations in helping to resolve the aforementioned small-scale problems are a natural and robust consequence of the physics implemented in the simulations, or is instead due to calibration (either implicitly or explicitly) of the feedback models to specifically address those issues, remains somewhat unclear\footnote{See \citet{Schaye2015} for a discussion of the predictive power of current cosmological hydrodynamical simulations.}.  In any case, the correct approach should not to be overly prescriptive about the nature of the cosmological model and the role of baryons, but to use comparisons of simulations to observations to attempt to constrain both.

At present, the literature contains numerous studies that typically explore one cosmological extension at a time, or examine the impact of baryons alone (in $\Lambda$CDM), and often in the context of a very small sample of haloes simulated (zooms).  Drawing general conclusions from these studies about the relative impact of these effects, their differences and  possible degeneracies, is hindered by large study-to-study variations in box size, resolution, differences in feedback calibration schemes, and focus on different aspects of small-scale structure.  Here we aim to remedy these issues by simultaneously studying the effects of the different extensions (SIDM, WDM, and a running spectral index) to $\Lambda$CDM in a consistent way, using a suite of high-resolution dark matter-only cosmological simulations.  
To compare to the impact of baryon physics, we make use of the high-resolution \texttt{EAGLE} Recal model \citep{Schaye2015,Crain2015} and the new \texttt{ARTEMIS} suite of zoomed hydrodynamical simulations of Milky Way-mass galaxies \citep{Font2020}. Note that \texttt{ARTEMIS} uses the same galaxy formation physics as \texttt{EAGLE} but the stellar feedback has been re-calibrated for the increased resolution to yield an improved match to the stellar masses of haloes with similar mass to the Milky Way.
Both of these suites have comparable resolution and statistics to our simulations of the above cosmological extensions.   
We examine the dark matter-only and hydrodynamical simulations in a consistent way, examining the impact of cosmological variations and baryon physics on a wide range of metrics that characterise the abundance and structure of dark matter haloes and their subhaloes.

The present study is structured as follows: In Section \ref{sec:sims} we introduce the simulations used as part of this study, covering each cosmological extension in turn. 
In Section \ref{sec:global_properties} we examine some global properties of the simulated volumes, focusing on statistics such as the halo and subhalo mass functions. 
In Section \ref{sec:internal_halo_properties} we examine the effects from these various cosmologies on several internal properties of simulated haloes, examining statistics such as the dark matter density profiles of haloes. Finally, in Section \ref{sec:conclusions} we summarise and discuss our findings.

\section{Simulations}
\label{sec:sims}
For this study, we have run a suite of high resolution, cosmological volumes, sampling different possible extensions to the $\Lambda$CDM model. 
The suite consists of 7 dark matter-only, periodic box simulations which are 25 comoving Mpc $h^{-1}$ on a side, each containing 1024$^3$ dark matter particles. 
All of the cosmological parameters for the simulations in this suite use the fiducial $Planck$ 2015 maximum likelihood values, apart from the simulations which include a running scalar spectral index (this is explained in more detail below). The cosmological parameter values adopted can be found in Table \ref{tab:cosmo_params}. 

The Boltzmann code \texttt{CAMB}\footnote{\href{http://camb.info/}{http://camb.info/}} (\citealt{Lewis2000}, August 2018 version) is used to compute the transfer functions and power spectra for all of the simulations (with the WDM power spectra being slightly altered, as explained below), and a modified version of \texttt{N-GenIC}\footnote{\href{https://github.com/sbird/S-GenIC}{https://github.com/sbird/S-GenIC}} is used to create the initial conditions (ICs) for the simulations.
The simulations are initialised at a starting redshift of $z$ = 127 and run to $z$ = 0. We use a modified version of \texttt{N-GenIC} which includes second-order Lagrangian Perturbation Theory corrections, alongside support for massive neutrinos.
Note that when producing the initial conditions (ICs) for the simulations, the same random phases are used for all 7 periodic box runs, removing the effects of cosmic variance when comparing the runs.

The simulations are run using a modified version of the parallel Lagrangian TreePM-SPH code \texttt{GADGET3} \citep[last described in][]{Springel2005a}. 

The gravitational softening is a fixed physical length of 250 pc $h^{-1}$ at $z\:\leq\:3$, and is a fixed comoving length at higher redshifts.
The reference $\Lambda$CDM simulation (designated `Ref') has a dark matter particle mass of $m_\textrm{DM}$ = 1.266 $\times$ $10^{6}$ M$_{\odot}$ $h^{-1}$. 
The SIDM and WDM simulations also have this particle mass, as they adopt identical cosmological parameters.  The two simulations which incorporate a running scalar spectral index have slightly different particle masses, owing to the slightly different values for $\Omega_\textrm{m}$ and $h$ for those runs, as described below. 
We note that for all the $Planck$ 2015 maximum-likelihood cosmologies used in this study, including the cosmologies with a running spectral index, we include massive neutrinos using the minimum summed neutrino mass (equal to $\Sigma M_{\nu}$ = 0.06 eV), derived from atmospheric and solar oscillation experiments and adopting a normal hierarchy of neutrino masses \citep{LesgourguesJ2006}.
As a result we use a version of \texttt{GADGET3} which has the semi-linear algorithm (developed by \citealt{Ali-Ha2013}, however see also \citealt{Bond1980, Ma1995, Brandbyge2008, Brandbyge2009, Bird2012}) which models the effects of massive neutrinos on both the background expansion rate and the growth of density fluctuations implemented into it. 
This algorithm computes neutrino perturbations on the fly at each time step (see \citealt{McCarthy2018} for further details of its implementation).  Furthermore, we also include the effects of radiation on the expansion history.

\subsection{Running of the scalar spectral index}

The epoch of inflation seeded small density perturbations in the matter distribution of the Universe, with the power spectrum of these perturbations in the $\Lambda$CDM model being of the form \citep{Guth1981}:

\begin{ceqn}
\begin{align}
\label{Pk_lambdacdm}
    P(k) = Ak^{n_s}
\end{align}
\end{ceqn}
\noindent where $P(k)$ is the power spectrum as a function of wavenumber $k$, $A$ defines the amplitude of the primordial matter power spectrum, and $n_s$ is the scalar spectral index.
In the $\Lambda$CDM model, $n_s$ is assumed to be constant, with no $k$ dependence.  However, as discussed in the Introduction, even the simplest models of inflation predict some level of deviation from a power-law distribution.

Allowing for running, the modified power spectrum, $P(k)$, can be expressed as \citep{Kosowsky1995}:
\begin{ceqn}
\begin{align}
    \label{eq:Pk}
    P_s(k) = A_s(k_0)\left(\frac{k}{k_0}\right)^{n_s(k=k_0)+\alpha_s'(k)}
\end{align}
\end{ceqn}
\noindent where $\alpha_s'(k)\equiv (\alpha_s/2)\ln(k/k_0)$, $\alpha_s$ is the running of the scalar spectral index, which is defined as d$n_s(k)$/d$\ln(k)$. 
The pivot scale, $k_0$, is the scale at which the amplitude of the power spectrum ($A_s$) is defined, along with the scale at which the spectral index is measured when $\alpha_s$ $\neq$ 0. In this study, we adopt the same pivot scale as that used for the cosmological parameter estimation of \cite{PlanckXIII}: $k_0$ = 0.05 Mpc$^{-1}$.  It is also worth mentioning that this is only a first-order extension to the power spectrum of density perturbations. There can also be a `running of the running', where the running of the scalar spectral index also varies with scale [$\alpha_s \rightarrow \alpha_s(k)$] leading to a second-order term being added to the functional form of the spectral index $n_s(k)$.  But for simplicity, we focus here on the first order effect.

As mentioned previously, all of the simulations, with the exception of the two simulations that include a running scalar spectral index, adopt the $Planck$ 2015 maximum likelihood cosmological parameters.
For the two simulations which include a running scalar spectral index, we adopt the same cosmological parameters as that derived in \citet{Stafford2020}. 
Here we will briefly explain how these cosmological parameters were chosen, but refer the reader to \cite{Stafford2020} for more detail.

To generate a set of parameters which make up a cosmological model, we made use of the $Planck$ 2015 publicly available Markov chains which include $\alpha_s$ as a free parameter.  
Five values for $\alpha_s$ which sampled the posterior distribution were chosen in \citet{Stafford2020}, corresponding to the maximum-likelihood value and $\pm1\sigma$ and $\pm2\sigma$ values.
For a given choice of the running, the values of the other important cosmological parameters were obtained by taking the weighted average of the Markov chain data for all parameter sets with $\alpha_s$ close to the desired value.  
Choosing the cosmological parameters in this way ensures that the resultant cosmological model is consistent with current measurements of the CMB.  In the present study, we use 2 out of the 5 cosmologies generated in this way, corresponding to the $\pm2\sigma$ values of the posterior distribution of $\alpha_s$.

\subsection{Warm dark matter}
Instead of modifying the primordial power spectrum, as in the case of a running of the scalar spectral index, the nature of dark matter itself can be modified.  Warm dark matter (WDM) differs from CDM in the standard model, in that the subatomic particles that constitute WDM are considerably lighter ($\sim$keV) than CDM particles ($\sim$GeV to TeV masses).  Consequently, WDM particles remain relativistic for longer in the early Universe compared with CDM, resulting in non-negligible thermal velocities that allow the particles to free-stream out of small-scale density perturbations, smoothing them out and suppressing the growth of structure on small scales \citep{Bond1983, Bardeen1986}.

This smoothing out of density perturbations due to the thermal velocity associated with WDM leads to a characteristic cut-off in the WDM power spectrum.  This effect can be modelled as a transfer function $T_{\textrm{WDM}}(k)$, relative to the CDM power spectrum:
\begin{ceqn}
\begin{align}
    P_{\textrm{WDM}}(k) = T_{\textrm{WDM}}^2(k)P_{\textrm{CDM}}(k).
\end{align}
\end{ceqn}

Here we make use of the fitting formula of \citet{Bode2001}:
\begin{ceqn}
\begin{align}
    T_{\textrm{WDM}}(k) = \left[1+(\alpha k)^{2\nu}\right]^{-5/\nu},
\end{align}
\end{ceqn}
\noindent with $\nu$ being a fitting constant and $\alpha$ corresponding to the scale of the cut-off in the power spectrum, with this being dependent on the mass of the thermal WDM particle.
Note that this function assumes the dark matter is entirely composed of WDM. The values adopted for these constants correspond to the best fit values obtained by \cite{Viel2005} (for $k-$scales < 5 Mpc$^{-1}\;h$); i.e., $\nu$ = 1.12 and (assuming the WDM is composed of thermal relics)
\begin{ceqn}
\begin{align}
\label{eq:alpha}
    \alpha = 0.049 \left(\frac{m_x}{1\textrm{keV}}\right)^{-1.11} \left(\frac{\Omega_x}{0.25}\right)^{0.11}
    \left(\frac{h}{0.7}\right)^{1.22}\textrm{Mpc}\:h^{-1}.
\end{align}
\end{ceqn}
Here $m_x$ corresponds to the mass of the WDM particle, $\Omega_x$ is the present-day density of WDM in units of the critical density, and $h$ is the reduced Hubble's constant. 

Examining equation \ref{eq:alpha}, it can be seen that the warmer the dark matter particle (i.e., the smaller the mass of the particle), the larger the scale of the break in the power spectrum (i.e., the suppression moves to smaller $k$ values).

As well as using a modified power spectrum as outlined previously, one should in principle also assign a thermal velocity to the WDM particles; with the rms velocity dispersion of these WDM particles being around 1.6 (0.6) km s$^{-1}$ in the case of the 2.5 (5.0) keV WDM particle mass at the starting redshift of 127 \citep{Bode2001}, which is a fraction of the rms velocity assigned through the Zel`dovich Approximation (around 40 km s$^{-1}$ at $z=127$).
Therefore, as the distance scales travelled during the free-streaming of the WDM particles due to their thermal velocities corresponds to $\sim$ tens of comoving kpc for WDM particle masses of $\sim$ keV \citep{Lovell2012}, we neglect these thermal velocities in our simulations. 
Furthermore, for a given initial power spectrum, the inclusion of thermal velocities tends to introduce spurious noise which can adversely affect early structure formation in the simulations \citep{Leo2017}.
Note that the particles are, however, still initialised with a peculiar velocity using the Zel`dovich Approximation \citep{Zeldovich1970} plus 2LPT corrections.

In this study we investigate two different dark matter particle masses: a 2.5 and 5.0 keV thermal relic dark matter mass. 
The choice for the WDM particle mass is influenced by current astrophysical constraints.   For example, a strong lower-limit is placed on the mass of the WDM particles through observations of the Lyman-$\alpha$ forest flux power spectrum, with \cite{Viel2013} placing a lower-limit of M$_{\textrm{WDM}} \gtrsim 3.3$ keV and \cite{Irsic2017} providing a slightly more stringent constraint of M$_{\textrm{WDM}} \gtrsim 5.3$ keV. 
These limits are consistent with constraints set from the Milky-Way's satellite population, which provide constraints ranging from M$_{\textrm{WDM}} > (1.5 - 3.9)$ keV \citep{Lovell2014, Kennedy2014, Jethwa2018, Nadler2019, Nadler2020b}.

Note, however, that the limits placed on the WDM particle mass through observations of the Lyman-$\alpha$ forest are somewhat sensitive to the treatment of the thermal history of the intergalactic medium (IGM). For example, \cite{Garzilli2019} demonstrated that the Lyman-$\alpha$ forest constraints can be lowered to $M_{\textrm{WDM}}>1.9$ keV when marginalising over the plausible temperature range of the IGM.

More recently, time delay measurements of (strong) gravitationally-lensed quasars, which are sensitive to the distribution of substructure around the lens, have been used to derive M$_{\textrm{WDM}} > 5.58$ keV \citep{Hsueh2019}.
As such, the thermal relic masses which we choose to simulate are consistent with the current tightest constraints placed on this parameter by various observations.

As this study is focused on dark matter (sub)haloes, it is important that the results are not influenced by numerical artefacts which may form. 
This is particularly relevant in the simulations of WDM universes, with some studies having shown an onset of structure formation on small scales in WDM simulations due to spurious fragmentation of filaments. \cite{Wang2007} showed this to be due to particle discreteness effects.  They provided an empirical cut in halo mass to remove these spurious objects from halo catalogues \citep[see also][]{Lovell2014}.  However, as this limiting mass is lower than the the limit we place on resolved (sub)haloes, such spurious structure formation should not impact our results (see Appendix \ref{sec:convergence_tests} for more details).

\subsection{Self-interacting dark matter}
Another possible extension to the $\Lambda$CDM model is to allow the dark matter particles to self-interact (i.e., scatter with neighbouring particles).  Here the dark matter is still assumed to be cold initially as in the standard model. 

To simulate the effects of these self-interactions, we use a slightly modified version of the previously mentioned TreePM \texttt{GADGET3} $N$-body code. This version was modified by \citet{Robertson2019} (see also \citealt{Robertson2017a,Robertson2017b}) to incorporate dark matter self-interactions. We will provide a brief overview of the interaction process here, but refer the reader to the original papers for more details. 

At each time-step $\Delta t$, particles search for any neighbouring particles that are within some pre-defined radius $h$. The probability for each pair of nearby particles $i$ and $j$ to scatter is calculated using:
\begin{ceqn}
\begin{align}
    P^{\textrm{scat}}_{ij} = \frac{|\textit{\textbf{v}}_i-\textit{\textbf{v}}_j| \Delta t}{\frac{4\pi}{3}h^3} . \left(\frac{\sigma}{m}\right)m_{\textrm{DM}}.
\end{align}
\end{ceqn}
Here ($\sigma/m$) is the particle physics dark matter scattering cross-section, and $m_{\textrm{DM}}$ is the dark matter particle mass in the simulation. $\textit{\textbf{v}}_{i,j}$ is the velocity of particle $i$ and $j$ respectively, and $h$ is the search radius which we set to be equal to the gravitational softening length $\epsilon$ of the simulations (see \citealt{Robertson2017a} for discussion).
Note this equation leads to the probability of a scattering event between two particles taking place being proportional to the dark matter particle mass.
As such, the total probability of a particle scattering with any other particle is independent of particle mass, as the number of neighbouring particles available to scatter with is inversely proportional to the dark matter particle mass.

The dark matter interactions are assumed to be fully described by an azimuthally-symmetric differential cross-section, defined in the centre of momentum frame of the two particles.
Although the code has the functionality for scattering events to be both velocity and angular dependent, in this study we only examine the case where the scattering events are velocity independent and isotropic.  Under these assumptions, the interactions are described by a differential cross-section (which is the rate at which particles are scattered into a region of solid angle) equal to: 
\begin{ceqn}
\begin{align}
\frac{d\sigma}{d\Omega} = \frac{\sigma}{4\pi}.
\end{align}
\end{ceqn}

We explore two different values for the total cross-section: $\sigma/m = 0.1$ and 1.0 cm$^2$ g$^{-1}$. 
These values are guided by current observational constraints.
Current constraints come from observations of strong lensing arcs, which suggest upper limits of $\sigma/m \la$(0.1, 1.0) cm$^2$ g$^{-1}$ \citep[][respectively]{Meneghetti2001, Robertson2019};
DM--galaxy offsets in colliding galaxy clusters, which suggest $\sigma/m <$(0.47 - 2) cm$^2$ g$^{-1}$ \citep{Randall2008, Kahlhoefer2015, Harvey2015, Kim2017, Robertson2017a, Wittman2018}; cluster shapes, have been used to infer $\sigma/m <$ (0.02, 0.1) cm$^2$ g$^{-1}$ \citep[][respectively]{MiraldaEscude2002, Peter2013}; 
as well as from subhalo evaporation arguments that suggest $\sigma/m <$ 0.3 cm$^2$ g$^{-1}$ \citep{Gnedin2001}. 
Note, however, that these constraints are mainly derived from galaxy cluster scales.  If the focus is turned instead to the dwarf-galaxy regime, cross-sections as high as 50 cm$^2$ g$^{-1}$ are within observational constraints and can potentially alleviate some of the small-scale problems associated with $\Lambda$CDM \citep{Elbert2015}.
As such, a velocity-dependent cross-section may be needed to explain the entire dynamic range of observations.
However, as mentioned, this study focuses on velocity-independent cross-sections, with the cross-sections chosen to be representative of the current constraints which exist for this parameter.

\subsection{Baryonic effects in the standard model}
\label{sec:EAGLE}

As discussed in the Introduction, many recent studies have concluded that the effects of baryons on `small-scale' structure are important.  
A primary aim of the present study is to compare and contrast the effects of baryons with those of changes to the standard cosmological model.  
We do this by comparing the gravity-only simulations introduced above to one set of simulations from the \texttt{EAGLE} project (the highest-resolution `Recal' run).  We also make comparisons to the \texttt{ARTEMIS} simulations, a new suite of fully hydrodynamic zoom-in simulations of Milky Way-type analogues \citep{Font2020}.
This new suite of zoom-in simulations uses the same galaxy formation (subgrid-)physics as the \texttt{EAGLE} project, but the stellar feedback has been re-calibrated for the higher resolution as well as to yield an improved match to observed galaxy stellar masses at halo masses of $\sim 10^{12} {\rm M}_\odot$.

Note that both sets of hydrodynamical simulations that we use fall within the context of the standard cosmological model.  Ideally, one would also like to explore the impact of baryons in non-standard cosmologies, since star formation and feedback-driven winds are observed to be ubiquitous in the Universe.  We leave this for future work, noting that it is not likely to simply be a matter of re-running the same galaxy formation model in each cosmological extension, as the properties of the galaxies themselves will likely change.  Therefore, re-calibration of feedback prescriptions will presumably be required.  In other words, some degree of degeneracy between baryonic effects and cosmological effects is to be expected.  We comment on this possibility below, using the simulations in hand.

\subsubsection{\texttt{EAGLE} simulation}

In this study we make use of the highest-resolution \texttt{EAGLE} run (data from \citealt{McAlpine2016}), called `Recal', which is a 25 Mpc on a side cosmological volume \citep{Schaye2015, Crain2015}. For a full overview of this and other \texttt{EAGLE} runs, including how the initial conditions were generated, we refer the reader to the aforementioned papers.  We provide only a brief overview of the Recal simulation here.

The simulation was run in the context of a $Planck$ 2013 maximum-likelihood cosmology \citep{Ade2014}, with the cosmological parameters provided in Table \ref{tab:cosmo_params}. 
The simulation consists of a fully hydrodynamic run, containing 2 $\times$ 752$^3$ particles, and a complimentary dark matter-only run (containing 752$^3$). This results in particle masses of $m_{\textrm{DM}}=1.21\times10^6$ M$_{\odot}$ and $m_{\textrm{gas}}=2.26\times10^5$ M$_{\odot}$.  The gravitational softening is 350 pc (in physical below $z=3$ and is a fixed comoving length at higher redshifts).  Thus, the mass and force resolution of this run is very similar to that of our dark matter-only suite introduced above.

The galaxy formation (subgrid-)physics included as part of the \texttt{EAGLE} project include metal-dependent radiative cooling \citep{Wiersma2009a}, star formation \citep{Schaye2008}, stellar evolution, mass-loss and chemical enrichment from Type II and Ia supernovae, Asymptotic Giant Branch and massive stars \citep{Wiersma2009b}, black hole formation and growth \citep{Springel2005b, Rosas-Guevara2015}, stellar feedback \citep{DallaVecchia2012} and feedback from active galactic nuclei (AGN) \citep{Booth2009}.

Like the larger (but lower resolution) Reference \texttt{EAGLE} run, the stellar feedback in the Recal model was adjusted to approximately reproduce the local galaxy stellar mass function and the size--stellar mass relation \citep[see][]{Schaye2015}.  In practice, the Recal model produces stellar masses that are somewhat below those inferred from the observed galaxy stellar mass function near the knee of the mass function (i.e., for Milky Way-mass halos, the stellar masses are somewhat too low), otherwise the galaxy stellar masses match the observations rather well.
We therefore compare the results derived from \texttt{EAGLE} with those derived from the \texttt{ARTEMIS}  zooms simulations (described below), for which the stellar masses agree better with observations for Milky Way-mass haloes.

\subsubsection{The \texttt{ARTEMIS} simulations}
\label{sec:ARTEMIS}

This study also makes use of the \texttt{ARTEMIS} simulations. This is a suite of zoom-in simulations, which are introduced in detail in \citet{Font2020} but we provide a brief overview here. 

The simulations were performed using a $WMAP9$ cosmology \citep{Hinshaw2013}, with the cosmological parameters for this run provided in Table \ref{tab:cosmo_params}.

Note that the $WMAP9$ maximum-likelihood cosmology does not include massive neutrinos, whereas, as mentioned previously, the $Planck$ 2015 cosmologies used for the cosmological volumes do.  However, this should not impact the comparisons made here, as we focus mainly on the relative effects of the cosmological extensions, as well as the relative effects of baryonic physics, compared to a CDM dark matter-only prediction. 
Furthermore, it was illustrated by \cite{Mummery2017} that the effects of massive neutrinos are separable from the effects of baryonic physics. 

The underlying galaxy formation (subgrid-)physics which is included in these full hydrodynamic zoom-in simulations is identical to that used in the \texttt{EAGLE} project (as described in Section \ref{sec:EAGLE}).
However, the parameters which characterise the efficiency of stellar feedback were adjusted to match the observed stellar mass--halo mass relation, as inferred from abundance matching.  The motivation for the adjustment was two-fold: i) the \texttt{EAGLE} Recal model predicted too low stellar masses on the scale of Milky Way-mass haloes (as mentioned above); and ii) the effectiveness of feedback in suppressing star formation tends to increase with increasing resolution for a fixed feedback model, thus the feedback efficiency in the higher-resolution \texttt{ARTEMIS}  simulations was reduced with respect to the \texttt{EAGLE} Recal model (see \citealt{Font2020} details).

The MW analogues were selected from a parent cosmological volume, 25 Mpc $h^{-1}$ on a side, containing 256$^3$ particles.
These MW analogues were then run at higher resolution, using the zoom-in technique \citep[see e.g][]{Bertschinger2001}, with the ICs being generated by \texttt{MUSIC} \citep{Hahn2011}.
The ICs are generated at a starting redshift of $z = 127$, with a transfer function computed using the Boltzmann code \texttt{CAMB}. Furthermore the ICs are generated including 2LPT corrections.

The MW type objects were selected based solely on a mass criterion, with any halo in the original dark matter-only cosmological volume which had a spherical-overdensity mass (defined in Section \ref{sec:HMF}) in the interval $11.903 \leq \log_{10}(M_{200, \textrm{crit}}/\textrm{M}_{\odot}) \leq 12.301$ being deemed a MW analogue. 
In total, this suite of zoom-in simulations contains 42 MW-analogues simulated at high resolution, with each hydrodynamic zoom-in having a complementary dark matter-only zoom-in simulation.
The resultant resolution of these simulations is a dark matter particle mass equal to $m_{\textrm{DM}} = 1.17\times10^{5}$ M$_{\odot}\:h^{-1}$, and initial baryonic particle mass equal to $m_{\textrm{gas}} = 2.23\times10^{4}$ M$_{\odot}\:h^{-1}$. 
The gravitational softening length for these zoom-in simulations is set to 125 pc $h^{-1}$ in physical coordinates for $z \leq 3$, and is changed to a constant comoving scale at earlier times.  Thus, \texttt{ARTEMIS}  is somewhat higher resolution than the simulations described previously.

\begin{table*}
\caption{The cosmological parameter values for the suite of simulations used in this study are presented here. The columns are as follows: (1) The labels for the different cosmological extensions simulated in this study, as well as the hydrodynamic simulations which we use. (2) Hubble's constant. (3) Present-day dark matter density in units of the critical density of the Universe. (4) Present day baryonic density in units of the critical density. (5) Spectral index. (6) Amplitude of the initial matter power spectrum at a pivot scale of 0.05 Mpc$^{-1}$ for the cosmological variations (corresponding to the $Planck$ pivot scale), and 0.002 Mpc$^{-1}$ for the \texttt{ARTEMIS} simulations (corresponding to a $WMAP9$ pivot scale). (7) linearly evolved present-day amplitude of the matter power spectrum on scales of 8 Mpc $h^{-1}$. Note that when producing the ICs for the simulations, we use the value for $A_s$, meaning that the ICs are normalised by the CMB. (8) Simulation DM particle mass.}
\begin{adjustbox}{width=\textwidth,center}
\begin{tabular}{cccccccc}
\hline
(1) & (2) & (3) & (4) & (5) & (6) & (7) & (8) \\ \hline
Cosmology & $H_0$ & $\Omega_{\textrm{DM}}$ & $\Omega_{\textrm{b}}$ & $n_s$ & $A_s$ & $\sigma_8$ & $m_{\textrm{DM}}$ \\
& (km s$^{-1}$ Mpc$^{-1}$) & & & & ($\times 10^{-9}$) & & ($\times 10^{6}$ M$_{\odot}\:h^{-1}$) \\ \hline
Ref ($\Lambda$CDM) & 67.31 & 0.264 & 0.049 & 0.966 & 2.199 & 0.830 & 1.063 \\
$\sigma/m$ = 0.1 cm${^2}$ g$^{-1}$ & 67.31 & 0.264 & 0.049 & 0.966 & 2.199 & 0.830 & 1.063 \\
$\sigma/m$ = 1.0 cm${^2}$ g$^{-1}$ & 67.31 & 0.264 & 0.049 & 0.966 & 2.199 & 0.830 & 1.063 \\
$M_{\textrm{WDM}}$ = 2.5 keV & 67.31 & 0.264 & 0.049 & 0.966 & 2.199 & 0.830 & 1.063 \\
$M_{\textrm{WDM}}$ = 5.0 keV & 67.31 & 0.264 & 0.049 & 0.966 & 2.199 & 0.830 & 1.063 \\
$\alpha_s$ = -0.02473 & 67.53 & 0.264 & 0.050 & 0.962 & 2.349 & 0.851 & 1.060 \\
$\alpha_s$ = 0.00791 & 66.95 & 0.268 & 0.049 & 0.965 & 2.147 & 0.821 & 1.079 \\
\texttt{ARTEMIS} & 70.00 & 0.233 & 0.046 & 0.972 & 2.410 & 0.821 & 0.118 \\
\texttt{EAGLE} & 67.77 & 0.259 & 0.048 & 0.961 & 3.097 & 0.829 & 0.821 \\ \hline
\end{tabular}
\label{tab:cosmo_params}
\end{adjustbox}
\end{table*}

\bigskip

\section{Global properties}
\label{sec:global_properties}
Here we present results for the effects that these different cosmologies have on the structure which forms across the different simulations.
We start by examining the effects on global properties of the simulated volume, including the halo and subhalo mass function, and the subhalo V$_{\textrm{max}}$ function.
\bigskip
\begin{figure}
    \centering
    \includegraphics[width=\columnwidth]{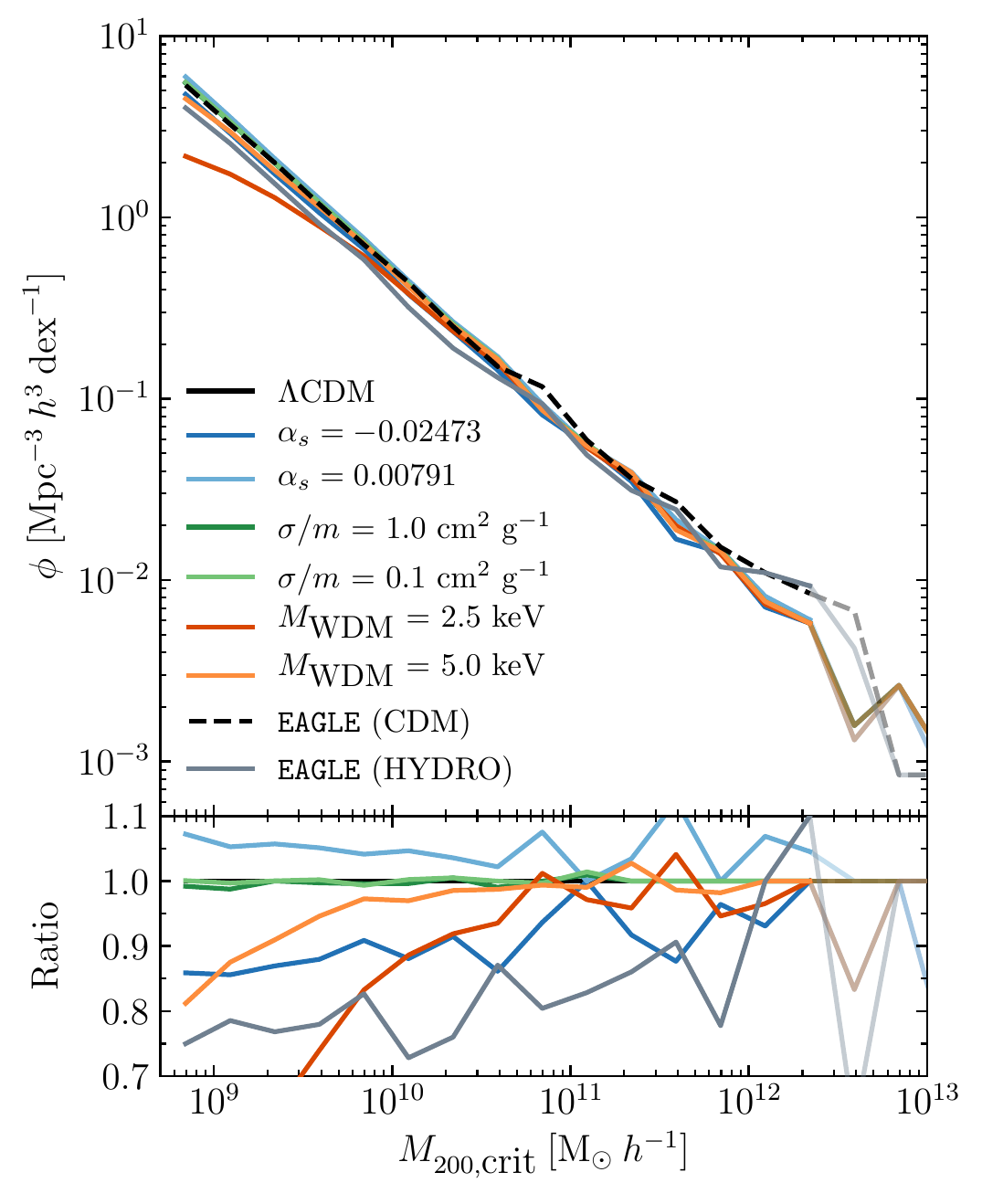}
    \vspace{-0.5cm}
    \caption{Top: number density of haloes per logarithmic mass interval, plotted as a function of their spherical overdensity mass. Bottom: the results in the top panel normalised with respect to the $\Lambda$CDM result. Note for the \texttt{EAGLE} simulations, the hydrodynamical simulation result is normalised with respect to the result from the complementary dark matter-only simulation. The lines become transparent for mass bins which have fewer than 10 haloes.}
    \label{fig:HMF}
\end{figure}

\subsection{Spherical-overdensity halo mass function}
\label{sec:HMF}
The first statistic we examine is the spherical-overdensity (SO) halo mass function (HMF), shown in Fig.~\ref{fig:HMF}.  Here we define the SO halo mass function as the number density of haloes which exist per logarithmic mass interval: $\phi \equiv dn/d\log_{10}(M)$, which in this study is plotted as a function of SO mass: $M=M_{200, \textrm{crit}}$. This is defined as the mass contained within a radius $R_{200, \textrm{crit}}$, which encompasses an overdensity 200 $\times$ the critical density of the universe.

Haloes in this study are identified using the \texttt{SUBFIND} algorithm \citep{Springel, Dolag2009}. First a standard Friends-of-Friends (FoF) algorithm is run on the dark matter distribution \citep{Davis1985}, linking all particles which have a separation less than some fraction of the mean inter-particle separation (with this fraction being set to 0.2 in this study).
With the FoF groups identified, \texttt{SUBFIND} then goes through each group and identifies locally bound substructures (subhaloes). 
The FoF group is centered on the position of the most-bound particle (lowest gravitational potential) in the central subhalo; i.e., the most massive subhalo in the FoF group. 
\texttt{SUBFIND} calculates a variety of subhalo parameters, including $M_{\textrm{sub}}$ (the summed mass of all particles deemed to be gravitationally-bound to a subhalo) and $V_{\textrm{max}}$ (the value at which a subhalo's circular velocity profile, $V(r) \equiv \sqrt{G M(<r)/r}$, reaches its maximum).  $V_{\textrm{max}}$ is a more stable quantity compared to the mass of a subhalo as it is less sensitive to how the subhaloes are identified and defined and it is more robust to tidal stripping \citep[e.g.][]{Diemand2007, Springel2008}. 

Focusing first on the cosmologies which have a running scalar spectral index as a free parameter, it can be seen in Fig.~\ref{fig:HMF} that there is a relatively large suppression ($\approx 10-20\%$) in the number density of haloes with $M_{200, \textrm{crit}} \la 10^{11}$ M$_{\odot}\:h^{-1}$ for the case with a negative running cosmology.   This is best illustrated in the bottom panel of this plot, which shows the HMF of each cosmology normalised with respect to the reference simulation\footnote{Note that the result shown for the \texttt{EAGLE} hydrodynamical simulation in the bottom panel is normalised by the complementary dark matter-only counterpart to the full hydrodynamical result. This is true for all ratio panels shown throughout this paper.}. 
Conversely the opposite is seen for the cosmology which has a positively-running scalar spectral index, which sees an $\approx 10\%$ increase in the number density of haloes at these masses. These results illustrate how relatively subtle changes to the primordial matter power spectrum are able to have relatively large effects on the growth of structure on small scales.

Similar to the effects of a negative running cosmology, switching to WDM also leads to a significant suppression in the mass function at low masses.  As expected, the effect is largest in the simulation with the lighter WDM particle mass. This is a well-known result, having been found for a range of WDM masses and models previously (e.g., \citealt{Smith2011, Angulo2013, Bose2016}).

These results illustrate the potential difficulty in constraining some of these cosmological extensions through simply counting the number of low-mass haloes. For example, the cosmology with a negatively-running spectral index can lead to a suppression very similar to that seen in the WDM cosmologies, particular at masses around 10$^9 \;\textrm{M}_{\odot}$.
Furthermore, if the cosmology happened to be a combination of either WDM + positive $\alpha_s$ or WDM + negative $\alpha_s$, this could lead to a universe very similar to a $\Lambda$CDM universe in the case of the former, and a universe with an apparently much warmer WDM model in the case of the latter.

Examining the simulations where the dark matter is allowed to interact with itself, there is no significant effect on the number density of SO haloes which form. This agrees with previous results that have been found when comparing large-scale statistics in SIDM and CDM cosmologies \citep[e.g.][]{Rocha2013}.
However, as we show below, SIDM can strongly affect the abundance of {\it satellite} haloes embedded within their hosts relative to that of the reference cosmology.  In addition, SIDM can have a large effect on the internal properties (e.g., density distribution), which we also discuss below. 

We now compare the previous effects, present due entirely to changes in the cosmological model, to those present in fully hydrodynamic simulations in the context of the standard model.  Examining the result from the \texttt{EAGLE} simulation, it is interesting that the suppression seen in the HMF is bracketed by the results for the two WDM cosmologies, and the cosmology with a negative running. 
For example the \texttt{EAGLE} simulation predicts a suppression of around 20\% for haloes with mass around 10$^{9}\;\textrm{M}_{\odot}\;h^{-1}$, whereas the suppression seen in the different cosmologies range between $\approx (15-50)\%$. Note that the suppression of the HMF for \texttt{EAGLE} was also shown previously by \cite{Schaller2015}.  Physically, the suppression of the halo mass function in the hydrodynamical simulations is the result of ejection of baryons due primarily to stellar feedback at these mass scales.  

Another interesting feature is how the suppression of the halo mass function as a function of mass in the \texttt{EAGLE} run is closely mimicked by the result seen for a (dark matter-only) cosmology with a negative running, at least in shape. Furthermore, as was shown in \cite{Stafford2020}, baryonic effects and the effects from the inclusion of a running spectral index appear to be separable to within a few percent, with each effect treatable as a multiplicative correction to the HMF of a $\Lambda$CDM gravity-only simulation.  As such, in the presence of an (unaccounted for) positively-running scalar spectral index, one would tend to underestimate the role of baryons in suppressing the HMF.
This problem is compounded by the current uncertainty in the suppression of the HMF due to baryonic physics on these scales. 
This is because current observations measuring the gaseous component of haloes on these scales \citep[e.g.][]{Giovanelli2005, Brown2017, Haynes2018} -- which are a proxy for how efficient stellar feedback is at removing baryons from these haloes -- are themselves quite uncertain. 
This propagates through to poor constraints on the stellar feedback implemented into hydrodynamical simulations. 
For example, as was shown in \cite{Davies2020}, there can be a large discrepancy in the gas fractions predicted at halo masses around $10^{11.5}\;\textrm{M}_{\odot}$ in simulations with different feedback implementations.
This was shown to be true when comparing the gas fractions seen in the \texttt{EAGLE} simulation and those seen in the \texttt{IllustrisTNG} simulation \citep{Nelson2018, Springel2018, Pillepich2018a, Pillepich2018b}. 
However, despite their differences, both of these simulations provide cold gas fractions which are consistent with present constraints \citep{Crain2017, Stevens2019}.

\begin{figure}
    \centering
    \includegraphics[width=\columnwidth]{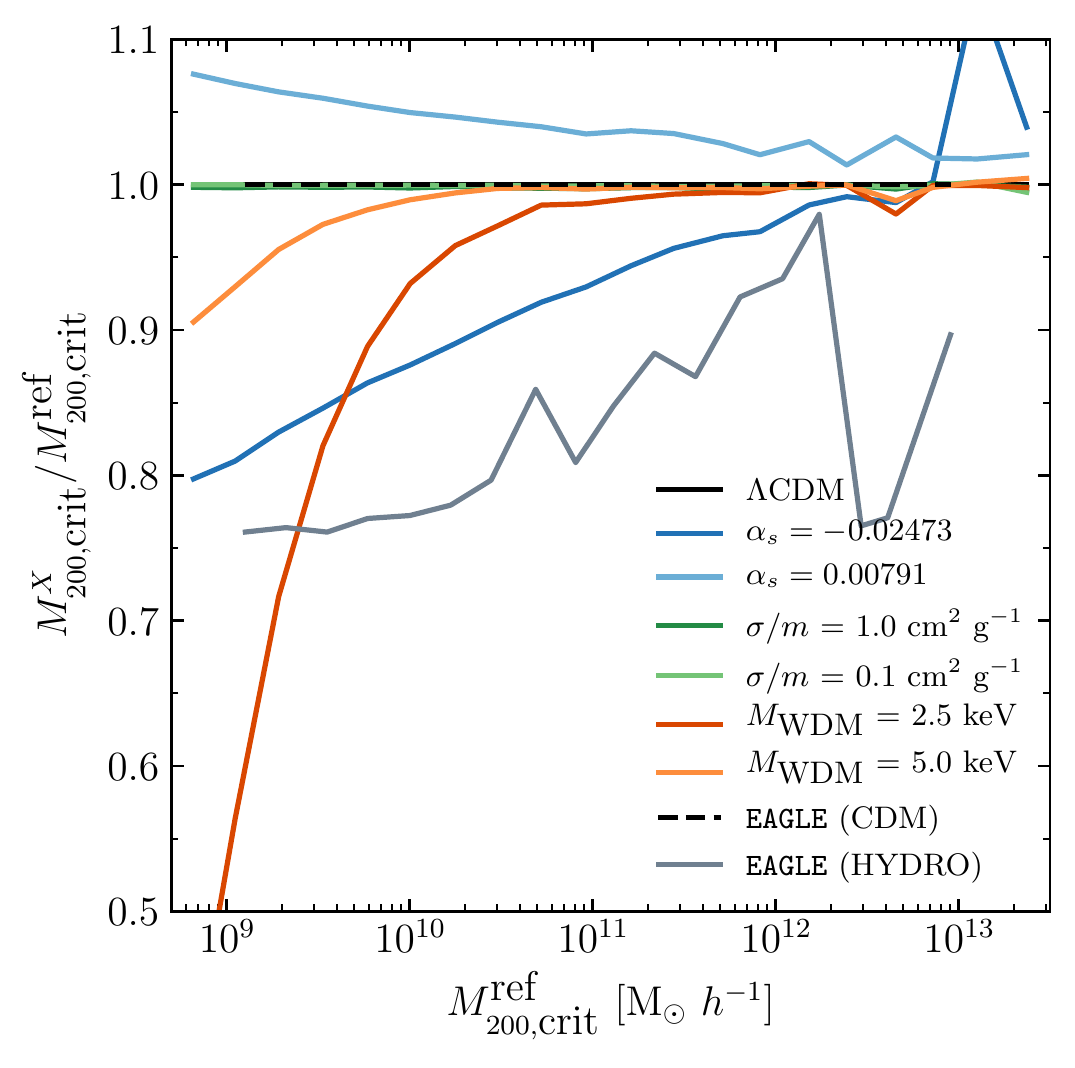}
    \vspace{-0.5cm}
    \caption{The fractional change in mass of a matched set of haloes across the different cosmologies, plotted as a function of the matched mass in the reference simulation, computed at $z=0$. This result provides an intuitive explanation for the effects seen on the HMF due to the changes in cosmology, or through the inclusion of galaxy formation physics in the simulations.}
    \label{fig:frac_mass_diff_haloes}
\end{figure}

The change in the SO halo mass function can be more readily understood by examining how individual haloes change their mass across the different cosmologies\footnote{This is possible since the simulations use the same initial phases.}. In order to calculate the change in mass, we first need to match haloes in different simulations using the unique IDs of the dark matter particles. We do this by bijectively matching the 50 most bound particles of haloes in the reference cosmology to each cosmological extension in turn. We only use haloes which successfully match both forwards and backwards in all simulations.

Using the list of matched haloes, we examine how the masses of individual haloes change across the different simulations.  Fig.~\ref{fig:frac_mass_diff_haloes} shows the median of the fractional change in mass of haloes in the different cosmologies, alongside the result for the \texttt{EAGLE} simulation. 

In terms of the effects of a running scalar spectral index, the positive-running cosmology boosts the halo masses by $\approx$5\% compared to its $\Lambda$CDM counterpart, whereas the masses are decreased in a negative-running cosmology by $>$ 10\% (with the largest effects coming at low masses). 
It is worth noting that, although the magnitude of this effect (and the effects seen throughout this paper on the various statistics examined) is larger in the negative running cosmology, this cosmology is also slightly more extreme than the cosmology with a positive running scalar spectral index (i.e. the initial density field has a larger difference with respect to $\Lambda$CDM in the case of the negative running cosmology). 
The reason for this is because the $Planck$ 2015 results slightly favour a negative running cosmology. 
As such, the effect on the initial matter distribution in this cosmology is larger than in the positive running cosmology (see e.g. fig. 2 in \citealt{Stafford2020}).
This effect echos what was found in \cite{Stafford2020}, extending the findings of that paper down to lower masses and illustrating further how the effects of a cosmology with a running spectral index are more pronounced on smaller scales.

There is also a strong effect on low-mass haloes in the most extreme WDM cosmology, with these haloes being less massive compared to their reference counterpart. 
This is expected due to the cut-off in the power spectrum on small scales, which preferentially affects low-mass haloes. The two different SIDM cosmologies have little effect on the halo mass over the entire mass-range sampled here.

Examining the results from the \texttt{EAGLE} simulation, again the result seen in the HMF is echoed here, with the mass of a matched set of haloes being reduced in the full hydrodynamic run, compared with its dark matter-only result.
The effect is strongest at lower masses, where stellar feedback in particular is able to efficiently blow gas out of the haloes, thus reducing their overall mass.  This, in turn, reduces the mass accretion rate onto the haloes, effectively amplifying the change in the final halo mass \citep{Sawala2013, Schaye2015}.
Note that the convergence of the halo mass in the \texttt{EAGLE} hydrodynamical simulation to the matched mass in the dark matter-only simulation, at around $M_{200, \rm{crit}}\approx10^{12}$ M$_{\odot}\;h^{-1}$, likely happens at too low of a mass.  Observations show that only haloes above $M_{200, \rm{crit}}\sim10^{14}$ M$_{\odot}\;h^{-1}$ are `baryonically closed' (e.g., \citealt{McCarthy2017}), whereas the \texttt{EAGLE} simulations demonstrate convergence to the universal baryon fraction at a considerably lower halo mass (see fig.~15 of \citealt{Schaye2015}).  A more realistic behaviour for the effects of baryons at high halo masses can be found in \citet{Velliscig2014} (see their fig.~2).

As already discussed, the similarity in the magnitude of the change to halo mass and, in some cases its dependence on mass itself (e.g., negative running cosmologies and baryonic effects have a similar mass dependence), strongly suggest that baryonic and cosmological effects on halo mass will be degenerate.  As we will show, however, the relative effects depend strongly on the nature of the probe, suggesting that multi-probe data sets are a potential way to break these degeneracies.

\subsection{The subhalo mass and \texorpdfstring{$V_{\textrm{max}}$}{Lg} functions}

We now examine the subhalo mass function (SMF) and subhalo maximum circular velocity function (SVF). Similarly to the HMF, these quantities are defined as the number of subhaloes of mass $M_{\textrm{sub}}$ (maximum circular velocity $V_{\textrm{max}}$) that exist per cubic comoving Mpc, per logarithmic mass (velocity) interval: $\phi \equiv dn/d\log_{10}(X)$ (where $X = M_{\textrm{sub}}$ or $V_{\textrm{max}}$).  As these quantities are well-defined for both central and satellite subhaloes (unlike SO masses), we investigate them separately.

\bigskip
\begin{figure}
    \centering
    \includegraphics[width=\columnwidth]{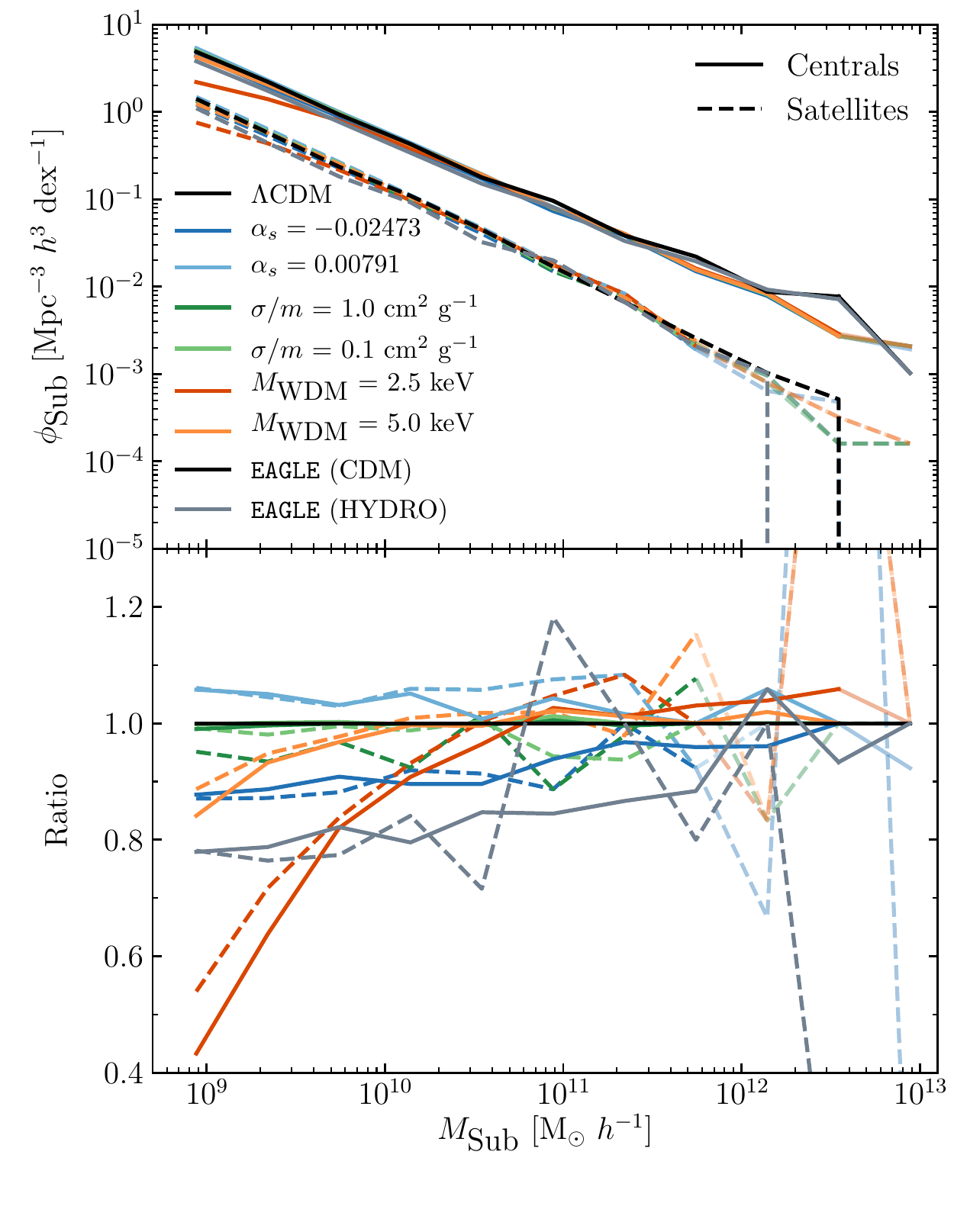}
    \vspace{-0.5cm}
    \caption{Top: the $z = 0$ number density of subhaloes as a functions of subhalo mass, split into the result for central subhaloes (solid) and satellite subhaloes (dashed). Note that this only includes subhaloes which have more than 200 particles bound to them, which was found to be a conservative resolution cut. Bottom: this shows the results in the top panel normalised to the result in the reference simulation, thus highlighting the differences between the simulations. The lines turn transparent for mass bins which have fewer than 10 subhaloes in. 
    }
    \label{fig:sub_mass_func}
\end{figure}

The SMF is shown in Fig.~\ref{fig:sub_mass_func}, which we split into the satellite and central mass functions.
A central subhalo is defined as the most massive subhalo contained within a single FoF group, with all other subhaloes in that same FoF group being defined as satellites. 
The bottom panel in this plot shows the result (be it the satellite or the central mass function) for each cosmology normalised with respect to the corresponding result in the reference $\Lambda$CDM simulation. 
Note that when examining subhaloes in this study, we focus on subhaloes with a mass $> 5\times10^{8}$ M$_{\odot}$, and $V_{\textrm{max}}>15$ km s$^{-1}$.
These values were calculated by comparing the $M_{\textrm{sub}}$-$V_{\textrm{max}}$ relation, along with the SMF, between two different resolution simulations.  This allows a conservative upper-limit to be placed on the values of $V_{\textrm{max}}$ and $M_{\textrm{sub}}$ for which these parameters are numerically converged (see Appendix \ref{sec:convergence_tests}).

\begin{figure}
    \centering
    \includegraphics[width=\columnwidth]{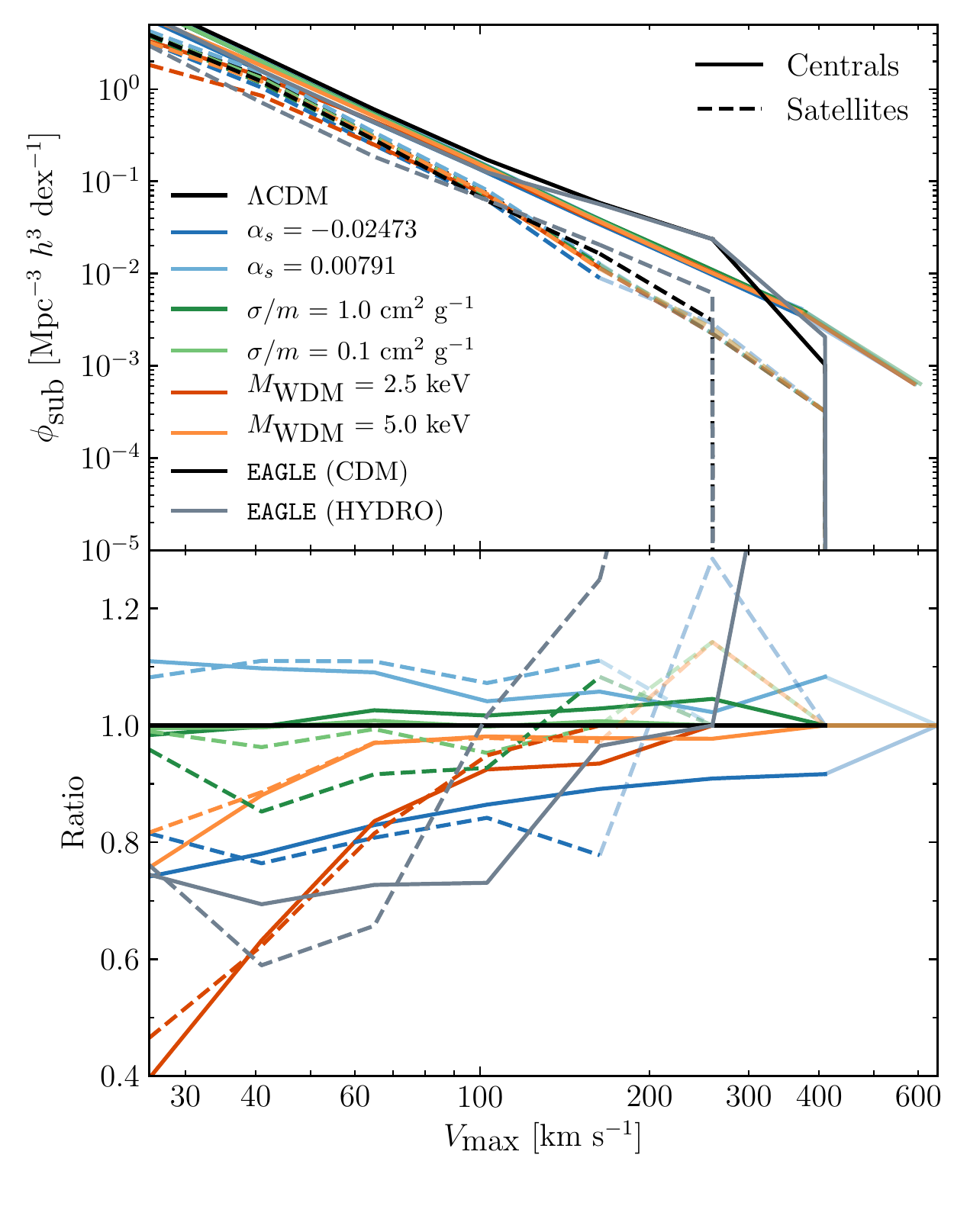}
    \vspace{-0.5cm}
    \caption{Top: the $z = 0$ number density of subhaloes, now as a function of $V_{\textrm{max}}$, once again split up into central subhaloes (solid line) and satellite subhaloes (dashed line) for the different cosmologies examined in this study. Bottom: the above results normalised with respect to the result in the reference cosmology, plotted to make the differences between the effects of each cosmology more apparent. Note the lines are made transparent for mass bins with fewer than 10 subhaloes present.}
    \label{fig:sub_Vmax_func}
\end{figure}

Focusing on the bottom panel of Fig.~\ref{fig:sub_mass_func}, the results largely mirror those for the SO HMF in Fig.~\ref{fig:HMF}. 
Furthermore, in general there are no large differences in the mass change for centrals and satellites for a given variation with respect to the reference $\Lambda$CDM model, with the exception of the SIDM simulations.
Here, in particular for the simulation with the higher cross-section, satellites appear to be more strongly affected than central subhaloes relative to the reference $\Lambda$CDM case.  We return to this point below.

Next we examine the subhalo V$_{\textrm{max}}$ function, which is shown in Fig.~\ref{fig:sub_Vmax_func}.  The effect that varying the underlying cosmology has on this is very similar to that seen in the SMF. 
For example, the WDM models and the negative running cosmology all lead to a large suppression in the number density of subhaloes at fixed $V_{\textrm{max}}$, with the positive running cosmology being the only model which produces an enhancement.  Similarly, in the most extreme SIDM cosmology, the number density of satellites is also suppressed, whereas there is little effect on the centrals. 
The result for the \texttt{EAGLE} simulation is also very similar to that seen in the SMF, in that at fixed $V_{\textrm{max}}$ there is a suppression in the number density of subhaloes in the hydrodynamic simulation. This is true for both centrals and satellites. 

Returning now to the difference between the impact of SIDM on satellites relative to central subhaloes, we believe this comes about due to the combination of two effects working in tandem.
Firstly, there is the `evaporation' of the subhaloes, which happens due to elastic collisions between the particles from a subhalo, and the particles of the host halo. These scattering events lead to neither particle being bound to the subhalo \citep{Vogelsberger2012}, causing the subhalo to lose mass as it infalls into the host halo. 
This evaporation mechanism was the original motivation for using SIDM to solve the missing satellites problem \citep{Spergel2000}.
However, another effect which causes the subhaloes to lose mass is tidal stripping. Due to the core which forms in subhaloes, they are also more susceptible to tidal stripping \citep{Penarrubia2010} (with \citealt{Dooley2016} finding this to be the more dominant effect out of the two). 
As such, both of these effects work to decrease the mass of a subhalo as it travels through its host halo, which is what drives the differences between the central and satellite mass ($V_{\textrm{max}}$) functions in the more extreme SIDM cosmology. 
We explore this effect further in Fig. \ref{fig:subhalo_rad_distrib}, which shows a strong suppression in the number density of satellite subhaloes in the central regions of their hosts in a SIDM cosmology, relative to the CDM case.

We have also examined the above statistics at early times, corresponding to $z = 0.5, 1.0$. However the results seen at these earlier redshifts are quantitatively very similar to the $z=0$ result, except that the results become more noisy at the high-mass end due to the finite box size.  For this reason, and for brevity, we only show the $z = 0$ results.

\subsection{Theoretical vs.~observable quantities}
\label{sec:theoretical_quants}

It is important to note that some of the aforementioned quantities are not strictly observable with current measurements.  For example, generally speaking, individual total halo masses can only be estimated with reasonable accuracy on the scale of galaxy groups and clusters\footnote{Measurements of total masses are possible for lower-mass systems (e.g., with galaxy-galaxy lensing), but generally require the stacking of large numbers of systems to make precise measurements.}, where a variety of methods exist for measuring masses (e.g., X-ray, gravitational lensing, and galaxy velocity dispersion measurements).  Thus, measurements of the halo mass function are currently very challenging on the scales we are interested in here.  On the other hand, observable quantities such as stellar mass are thought to be very good proxies for total halo mass, with a typical scatter in the stellar mass at fixed halo mass of 0.2 dex (e.g., \citealt{Behroozi2013}).  In terms of the quantity $V_{\rm max}$, while rotation curves are directly accessible observationally, current observations of low-mass galaxies struggle to reach the flat part of the rotation curve, so that estimates of $V_{\rm max}$ can sometimes require significant extrapolation.  

Ultimately, in order to make definitive statements about the degeneracies between baryon physics and changes in cosmology for low-mass galaxies, one should make synthetic (or `mock') observations from the simulations and analyse them in a way that is faithful to what is done for genuine observations.  In order to do this, however, a simulation suite which includes baryon physics for all cosmological variations is required.  We leave this for future work.

\section{Internal halo properties}
\label{sec:internal_halo_properties}

We turn now to the internal structure of host haloes across the different simulations. 
We stack haloes in 6 equally spaced logarithmic mass bins between $10.0 \leq \log_{10}(M_{200,\textrm{crit}}\;[\textrm{M}_{\odot}]) \leq 13.5$, with the results shown corresponding to the median result in each mass bin. 
In particular, we examine the spherically-averaged dark matter density profiles of haloes (which we also use to examine the concentration--mass relation later), along with the circular velocity profiles of the haloes.
We also examine how subhaloes are radially distributed in their host haloes.

\subsection{Dark matter density and circular velocity profiles}
\label{sec:density_profiles}
\begin{figure*}
    \centering
    \includegraphics[width=\textwidth]{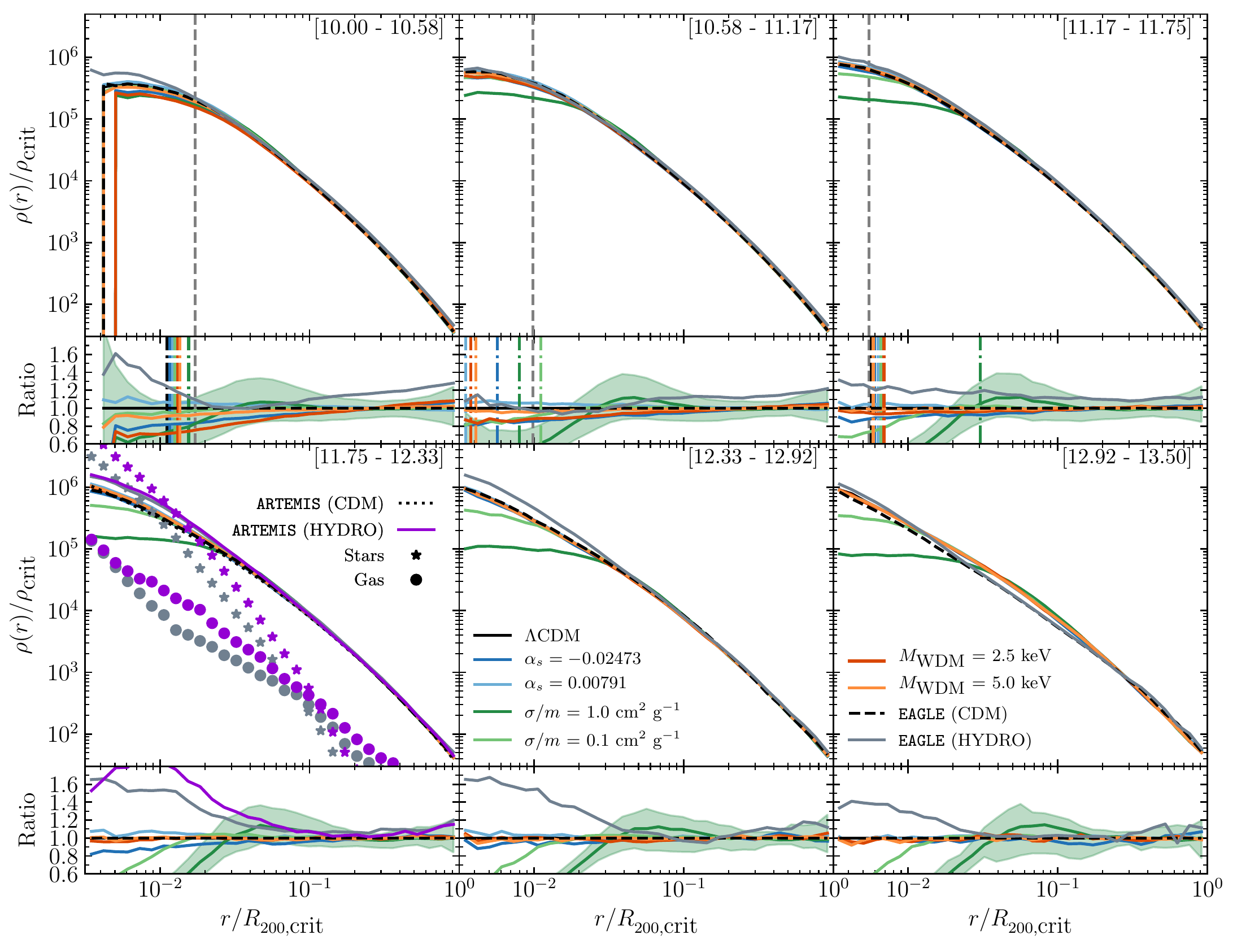}
    \vspace{-0.5cm}
    \caption{Spherically-averaged dark matter density profiles for the different cosmological models at $z=0$. The different panels here correspond to different mass bins, with the mass window being shown in the top-right hand corner of each panel (equal to $m_1 \leq \log_{10}(M_{200,\textrm{crit}}\;[\textrm{M}_{\odot}]) < m_2$). The different curves correspond to the median result in that mass bin. The smaller panels below the density profiles show the median density profiles in each mass bin normalised with respect to the median density profile in the reference cosmology in that mass bin. The result is also plotted for haloes in the \texttt{EAGLE} simulation. Alongside this, we also compare these results to the density profiles calculated for the MW analogues from the \texttt{ARTEMIS} simulations. The results for this is shown in the bottom-left panel. We also plot in this panel the stellar (shown by stars) and gaseous (shown by circles) density profiles for the two hydrodynamic simulations. The shaded regions show the 16th-84th percentile result for the most extreme SIDM cosmology. This scatter is representative of the other 6 cosmologies, and shows the significance of the core produced in the SIDM cosmologies' density profiles. The vertical dashed line represents the median convergence radius for that mass bin, calculated in the $\Lambda$CDM cosmology. The vertical dot-dashed lined shown in the ratio panel corresponds to the convergence radius calculated using the ratios as a convergence diagnostic rather than the actual density profiles themselves, as described in Section \ref{sec:conv_ratios}. Note that this radius corresponds to a conservative upper-limit on the convergence of the ratios, as it is calculated based on the minimum radius down to which a lower resolution simulation is converged with the simulations for which these results were extracted. We only plot this result for the first three mass bins, as it is for these mass bins that the absolute profiles are not converged over the entire radial range examined in this study. For instances where this convergence radius is exactly equal for different cosmologies we slightly displace them to larger radii for clarity.}
    \label{fig:density_profiles_box}
\end{figure*}

\begin{figure*}
    \centering
    \includegraphics[width=\textwidth]{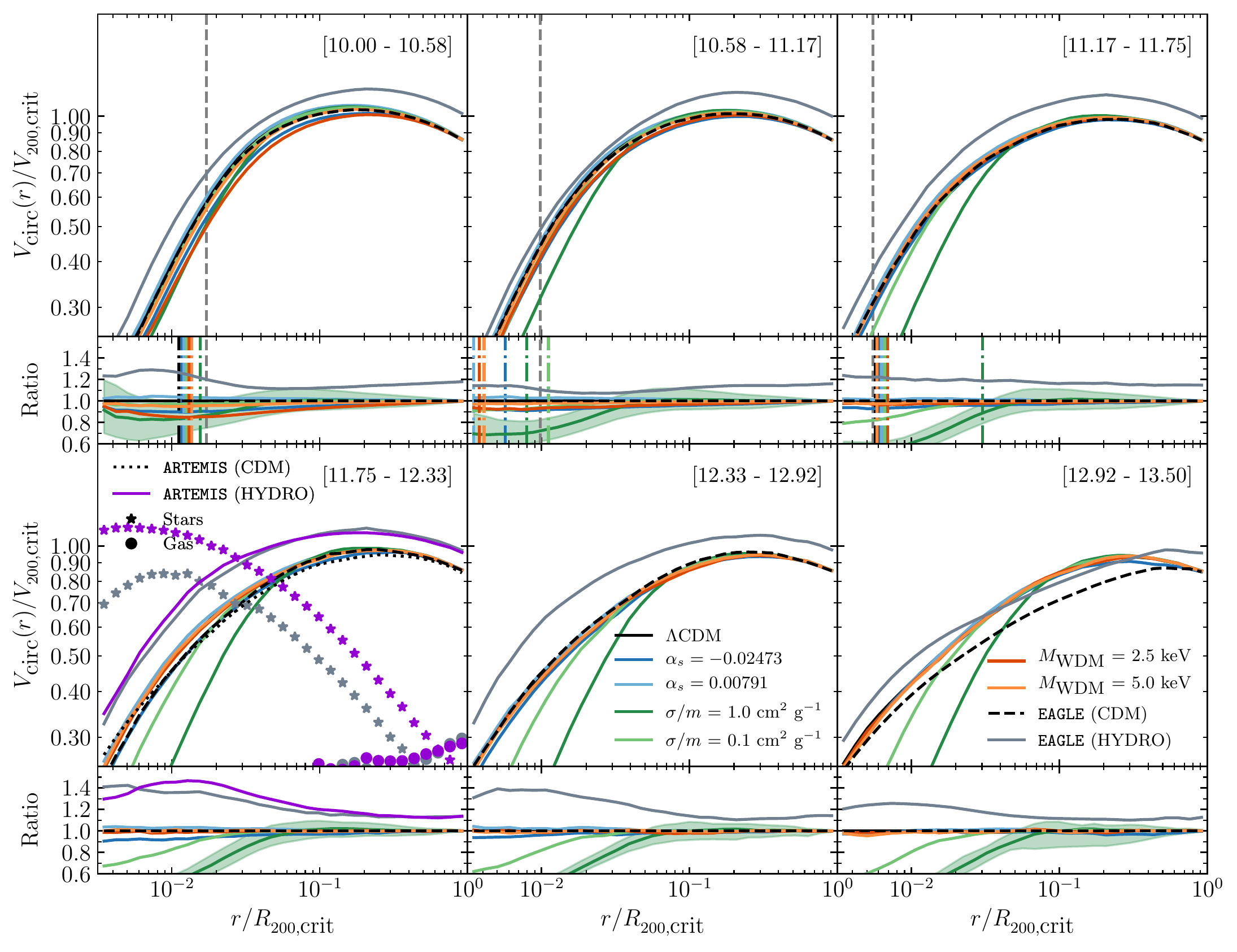}
    \vspace{-0.5cm}
    \caption{The median dark matter circular velocity profile of haloes in 6 separate mass bins for the different cosmologies examined in this study at $z = 0$, scaled by the virial circular velocity. 
    The result is plotted in the same mass bins examined in the previous plot, indicated in the top right-hand corner of each panel. 
    The bottom panel(s) below each main plot show the above result normalised with respect to the median result in that mass bin in the reference cosmology.
    Once again, the shaded regions correspond to the 16th-84th percentile plotted for the more extreme SIDM cosmology, with this scatter being representative of the other cosmologies.
    The vertical dashed line corresponds to the median convergence radius of haloes in that mass bin in the $\Lambda$CDM cosmology.
    The vertical dot-dashed line shown in the ratio panels corresponds to the convergence radius obtained from the Section \ref{sec:conv_ratios} analysis. Once again, we only show this result for the top three panels, and slightly displace any radii which are exactly equal across different cosmologies for clarity.}
    \label{fig:circ_vel_profile}
\end{figure*}

The spherically-averaged dark matter density and circular velocity profiles for all of the simulations we consider are shown in Figs.~\ref{fig:density_profiles_box} and \ref{fig:circ_vel_profile}, respectively.  To place the `dark matter-only' simulations on an equal footing the hydrodynamical simulations, we rescale dark matter particle masses in the dark matter-only simulations by the ratio $\Omega_{\textrm{CDM}}/\Omega_{\textrm{m}}$ (where $\Omega_{\textrm{m}}$ is the total matter density in units of the critical density).
As mentioned, the results shown in this section correspond to the median result of all haloes in that mass bin, with the mass bin indicated in the top right-hand corner of each panel.  The results are shown in dimensionless units, with the density profiles being scaled by the critical density of the universe at $z=0$ and the circular velocities scaled by the circular velocity at $R_{200, \rm{crit}}$, $V_{200, \textrm{crit}} = \sqrt{G M_{200, \textrm{crit}} / R_{200, \textrm{crit}}}$.
The profiles themselves are computed in 32 logarithmically-spaced radial bins between: $-2.5 \leq \log_{10}(r/R_{200,\textrm{crit}}) \leq 0$. 

The plots are split into 6 panels according to halo mass bin.  The smaller panels below the main panels correspond to the median density/circular velocity profile in that mass bin normalised with respect to the median density/circular profile in the reference $\Lambda$CDM cosmology in the same mass bin. 
The vertical dashed line in each panel corresponds to the median convergence radius calculated for the $\Lambda$CDM simulation.  We compute the convergence radius following that advocated by \cite{Ludlow2019} (see their equation 15). This is a follow-up study to that done in \cite{Power2003}, where the convergence radius is calculated explicitly for stacks of haloes.  Ludlow et al.~find that the median result of a stack of haloes (circular velocity profiles in their study) is converged to smaller radii than what was advocated by \cite{Power2003}.  
In this study, however, we are less focused on the absolute value of the density and circular velocity profiles and more on the \textit{relative} effects (shown by the ratio panels in each plot) on the profiles due to changes in cosmology and inclusion of baryons.  Therefore, as an aside, we also investigate below (see Section \ref{sec:conv_ratios}) whether the \textit{ratio} of the density profiles of haloes in non-standard cosmologies with respect to $\Lambda$CDM are potentially converged to smaller radii than that advocated by \cite{Ludlow2019}. 
This is particularly relevant for the first three mass bins, which are not converged over the entire radial range which we examine here.   We indeed find the ratios are generally converged to small radii and as such we also plot on Figs. \ref{fig:density_profiles_box}, \ref{fig:circ_vel_profile} the convergence radii that we advocate for these first three mass bins. 

The strongest cosmological effect on the profiles come from the SIDM model, particularly the run with the largest cross-section. The scattering acts as an efficient mechanism for dynamically heating and redistributing the mass in the inner regions, producing a near constant-density core. 
The size of the core relative to the virial radius is an increasing function of halo mass, this is expected given the fact that the scattering rate depends on both the local density and velocity. 
This means that the scattering rate at a fixed fraction of $R_{200, \textrm{crit}}$, scales with $V_{200, \textrm{crit}}$; as such, more massive haloes (higher $V_{200, \textrm{crit}}$) have higher scattering rates, at a certain fraction of $R_{200, \textrm{crit}}$, resulting in larger cores forming.
This effect has been seen in previous studies \citep[e.g.][]{Colin2002, Rocha2013, Vogelsberger2019, Robertson2019}.  Interestingly, in the $\sigma/m = 1.0$ cm$^2$ g$^{-1}$ cosmology, the redistribution of matter from the inner regions to the outer regions tends to lead to a small {\it enhancement} in the dark matter density on scales of $\approx0.05$--0.1 $R_{200,\textrm{crit}}$ relative to the baseline collisionless $\Lambda$CDM model.

The next largest cosmological effect is seen in the cosmology with a negative running of the spectral index. Here there is also a suppression in density in the inner-regions, particularly for the lower-mass haloes, with there being an $\approx$ 20\% decrease in the central density of these objects ($M_{200, \textrm{crit}}$ < 10$^{11}$ M$_{\odot}$). 
This effect decreases with increasing mass, with the profile in the highest-mass bin being broadly consistent with the result in the reference cosmology. 

This trend, however, continues further, as was shown in \cite{Stafford2020}; a cosmology with a negative running actually predicts an enhancement in the density profile of haloes with mass $\gtrsim 10^{14}\;\textrm{M}_{\odot}$.
Conversely to the SIDM models where there is a very sharp drop in the density profile and a small increase in density over a narrow radial range just outside of the core, the decrease in the density profile towards the center in the negative running cosmology is far more gradual.

There is also a slight decrease in the profiles in the inner regions in the two WDM models, particularly at the low mass end, with this effect disappearing quickly with increasing mass; with haloes of mass $M_{200, \textrm{crit}} \geq 10^{11.17}$ M$_{\odot}$ having density profiles almost indistinguishable from the $\Lambda$CDM result.
This agrees with the results of \cite{Bose2016}, along with the lower-concentration of low-mass haloes found in \cite{Ludlow2016} (see also Fig.~\ref{fig:conc_mass}).

The cosmology with a positive running of the spectral index is the only cosmology which appears to predict an enhancement in the central density (and circular velocity) profiles of haloes. However, this enhancement is generally quite mild and decreases with increasing halo mass.

We now investigate the impact of baryons on the dark matter density and circular velocity profiles, using the \texttt{EAGLE} and \texttt{ARTEMIS} simulations. 
For the three lowest-mass bins, the impact of baryons on the dark matter density profiles in the \texttt{EAGLE} simulations is relatively mild.  As the stellar mass fractions are very low in this regime \citep{Schaye2015, Schaller2015}, 
there is no significant adiabatic contraction of the dark matter.  On the other hand, \texttt{EAGLE} may underestimate the effects of repetitive feedback episodes in introducing a core in the dark matter, due to its lack of an explicit cold phase of the ISM and a relatively low density threshold for star formation (e.g., \citealt{BenitezLlambay2019}).
However, when examining mass bins with $M_{200, \textrm{crit}} \geq 10^{11.75}$ M$_{\odot}$, where it is not thought that feedback can produce such cores (e.g., \citealt{Dutton2011}), the inner density and circular velocity profiles appear far more pronounced in the hydrodynamic result compared to the dark matter-only result. 
This illustrates the contraction of the dark matter in the central regions due to the presence of stars in the inner regions of the central galaxies of these haloes. 
A similar result was shown by \cite{Schaller2015} (see their fig. 6), who also showed the density profile of stars in their host haloes, illustrating how these dominate the total matter density in these regions, particularly for these higher mass haloes.

Focusing on haloes in the mass range $10^{11.75} \leq M_{200, \textrm{crit}}/\textrm{M}_{\odot} < 10^{12.33}$ (shown in the bottom-left panel of the same figure), we also plot the median result for the dark matter density and circular velocity profiles computed for the 42 MW analogues from the \texttt{ARTEMIS} simulations.
Recall that the mass-range sampled by these zoom-in simulations is from: $10^{11.9} \leq M_{200, \textrm{crit}}/\textrm{M}_{\odot} < 10^{12.3}$, roughly corresponding to the same range examined for the cosmological extensions \footnote{When taking the ratio for the \texttt{ARTEMIS}, we divide the median result for the haloes in the full hydrodynamic run by the median result from the complimentary dark matter-only zoom simulations}. 
Similarly to the results seen in the \texttt{EAGLE} simulation, at large $r$, the dark matter density and circular velocity profiles in the hydrodynamic simulations are quite similar to the result seen for the dark matter-only simulation.
However, the density and circular velocity profiles in the inner regions of the haloes ($r \lesssim 0.1R_{200, \textrm{crit}}$) are far steeper in the \texttt{ARTEMIS} simulations than their dark matter-only counterpart.  This effect is also larger than that seen in the \texttt{EAGLE} simulations and extends to larger radii.  We argue that this is the case because \texttt{ARTEMIS} has higher stellar masses than \texttt{EAGLE} for this halo mass range (with the former being in better agreement with observations).

To demonstrate this, we also plot in this panel the density profiles of the stellar and gaseous components of the halo. Here one can see that \texttt{ARTEMIS} has a systematically larger stellar density at all radii compared to the \texttt{EAGLE} result (along with an increased circular velocity). It is these differences which produce the differing profiles between the two hydrodynamical simulations. 
Note that the apparent offset between the density and circular velocity profiles in the hydrodynamic simulations relative to the gravity-only simulations is due to the former having a larger median mass in each of the mass bins.

These results demonstrate the importance of having simulations which also include a prescription for galaxy formation physics, as baryonic physics has one of the largest effects on the underlying dark matter distribution in Figs.~\ref{fig:density_profiles_box} and \ref{fig:circ_vel_profile}. 
This is particularly apparent when examining the panels which also show the stellar density (circular velocity) profiles. These show how the stellar component dominates the profiles in the inner regions and, via adiabatic contraction, produces the changes seen in the dark matter profiles. It is in these regions also where the different cosmological models are seen to have the largest effects.
Therefore, as discussed in the Introduction, the ideal scenario is to simulate not only the standard cosmological model, but also plausible non-standard cosmologies using full cosmological hydrodynamical simulations.  
To date, only a small number of studies have attempted this.  For example, recent simulations presented in \cite{Despali2019} explored SIDM hydrodynamical simulations, concluding that the gravitational potential of the stars at the centre of intermediate mass haloes ($M_{\textrm{vir}}\:\approx\:10^{12}\;\textrm{M}_{\odot}$) significantly alters the DM distribution, removing the SIDM-induced core, while cores still formed in more massive haloes ($M_{\textrm{vir}}\:\approx\:10^{13}\;\textrm{M}_{\odot}$), with this being related to the formation time of the halo.
Using zoomed hydrodynamical simulations of clusters, \citet{Robertson2018} found a very complex interplay between baryons and SIDM, with some haloes showing dark matter cores, and others which do not.  
How the results of combined baryons + non-standard cosmology runs depend on the method for implementing and calibrating the feedback, the modelling of the cold ISM (or lack thereof), and the resolution of the simulations, has not yet been explored in much detail, but it would not be surprising if the conclusions about the interplay of baryons and cosmological effects were sensitive to these factors and should be explored.

The above analysis of the density and circular velocity profiles was done in bins of halo mass, using the self-consistently computed halo masses from each of the simulations.  Since the halo mass itself changes as a result of cosmological and baryonic effects (as we showed in Section \ref{sec:global_properties}), this implies that, in general, for a given mass bin we are not analysing precisely the same haloes in each simulation.  However, we have also examined a versions of Figs.~\ref{fig:density_profiles_box} and \ref{fig:circ_vel_profile} where we use matched haloes and binned the profiles according to the mass from the matched reference $\Lambda$CDM model.  In general, the effects are very similar but are slightly more pronounced in the matched halo plot.  For brevity, we do not include these figures here.

\subsection{Convergence radius of relative cosmological effects}
\label{sec:conv_ratios}
\begin{figure*}
    \centering
    \includegraphics[scale=0.7]{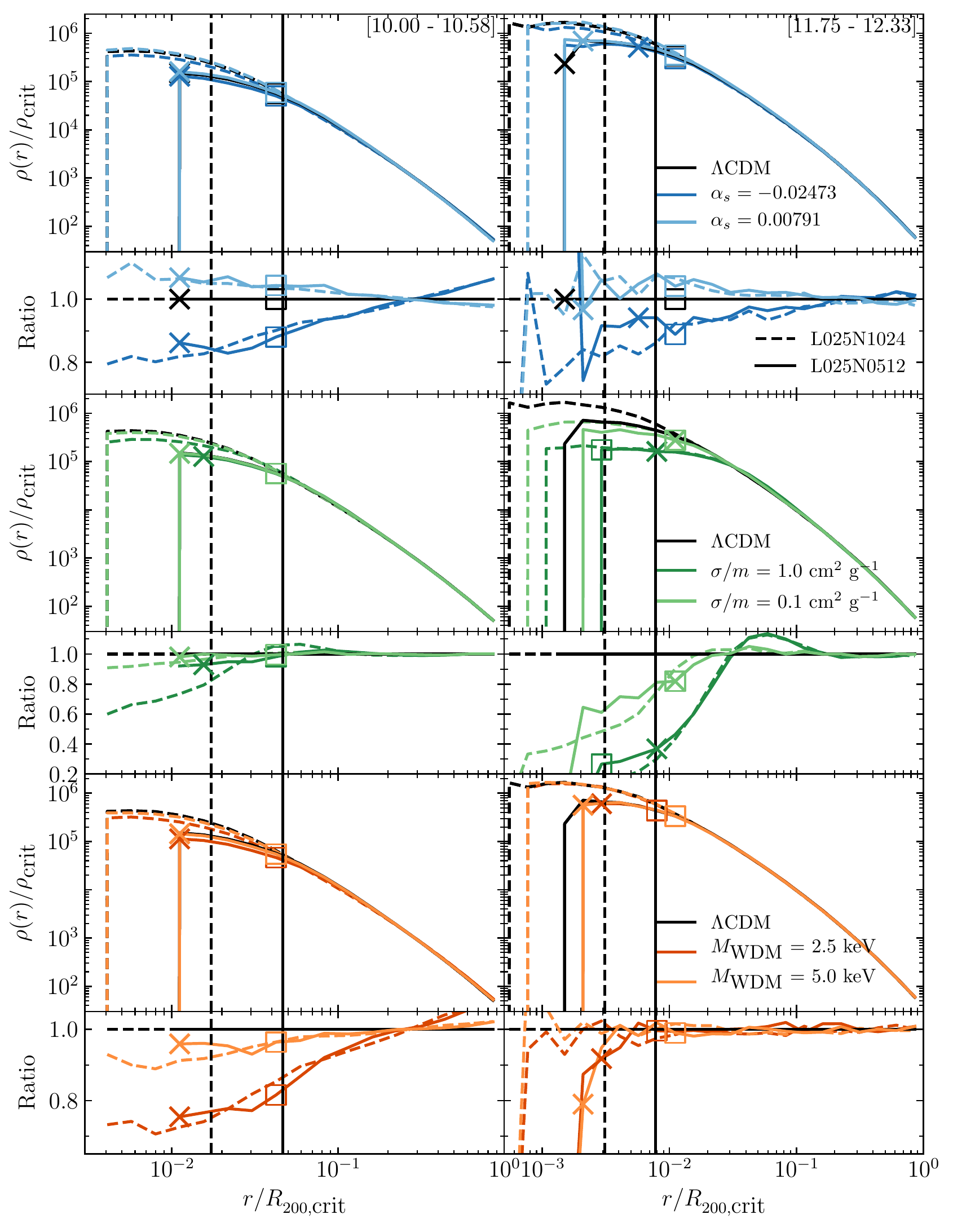}
    \caption{Comparison of the convergence of median spherically-averaged dark matter density profiles to the convergence of their ratios with respect to a $\Lambda$CDM cosmology. Columns are separated into different mass bins: Left are haloes in the mass range $10\leq\log_{10}\left(M_{200, \textrm{crit}}\;[\textrm{M}_{\odot}] \right) < 10.58$, right are haloes in the mass range $11.75\leq\log_{10}\left(M_{200, \textrm{crit}}\;[\textrm{M}_{\odot}] \right) < 12.33$. The different rows show the results for different cosmological extensions: top -- running spectral index, middle -- SIDM, bottom -- WDM. Panels below the absolute density profiles show the ratios with respect to the result in the $\Lambda$CDM cosmology. Crosses show the radius at which the ratios diverge by greater than 10\% relative to the higher resolution simulation. Hollow squares represent the radius at which the absolute density profiles diverge by greater than 10\%. The veritcal solid and dashed lines show the median \citet{Ludlow2019} convergence radius for the $\Lambda$CDM cosmology in that mass bin, for the different tiers of resolution.}
    \label{fig:conv_test_ratios}
\end{figure*}

There have been multiple studies which examined the convergence of internal halo properties in numerical simulations \citep[e.g.][]{Power2003, Diemand2004, Navarro2010, Ludlow2019}, which have shown that 2-body relaxation effects impose a lower-limit on the radius to which one can trust mass profiles of haloes. This was most recently investigated by \cite{Ludlow2019}, who provided a criterion on the number of particles needed below a radius $r$ for the median mass profiles of a stack of haloes to be converged. 

These studies have been extremely useful when interpreting gravity-only N-body simulations.  It is noteworthy, however, that they have all been done in the context of a $\Lambda$CDM universe and have focused on the convergence of the absolute values for the statistics examined.
In this study, we are instead mostly interested in the relative effects on certain statistics. In particular the ratio of, for example, the median spherically-averaged density profiles of haloes in non-standard cosmologies with respect to the result in the $\Lambda$CDM case.
For this reason, as an aside, we briefly explore the convergence of relative cosmological effects.  Readers who are less interested in the convergence properties may wish to skip ahead to Section \ref{sec:rad_sub_haloes}.

To examine the convergence properties, we have run an additional set of simulations but at lower resolution, with 512$^3$ particles, $m_{\textrm{DM},\Lambda \textrm{CDM}}=1.013\times10^{7}$ $\textrm{M}_{\odot}$ $h^{-1}$, and $\epsilon=500$ pc $h^{-1}$. We compute the density profiles in 32 logarithmically-spaced radial bins, ranging from: $-4.5 \leq \log_{10}\left(r/R_{200, \textrm{crit}}\right) \leq 0$ (note that this minimum radius is smaller than the minimum radius used in the previous section, this was chosen to test the ratios to as small a radius as possible).

The results are shown in Fig. \ref{fig:conv_test_ratios}, which is set up as follows: the columns represent two different halo mass bins (indicated in the bottom right corner of the top panel), the left column corresponds to the smallest mass bin of main FoF haloes examined in this study.
The right column corresponds roughly to a MW analogue mass window.  We look at these mass bins because, in the case of the former, these haloes will be most subject to numerical effects, and, in the case of the latter, because much of the emphasis of testing models of dark matter have been at MW scales (for obvious reasons). 
The rows in this plot correspond to the three different cosmological extensions (running, SIDM, WDM), indicated in the top-right of each right-column panel. 
We show both the median spherically-averaged dark matter density profiles for each cosmology, as well as, below this, the ratios of the median result taken with respect to the $\Lambda$CDM result.
We also plot the median convergence radius calculated using equation 14 from \cite{Ludlow2019} for the two different $\Lambda$CDM simulations, with their line-styles indicating their resolution. 

Similarly to what was done in previous convergence test studies, we define the convergence radius as the minimum radius for which $\left(R_{\textrm{high}}-R_{\textrm{low}}\right)/R_{\textrm{low}} \leq 0.1$ is true, where $R_{\textrm{high}}$ ($R_{\textrm{low}}$) is the ratio in the high-resolution (low-resolution) simulation of the density profile of a cosmological extension with respect to the $\Lambda$CDM result. This radius is shown by the coloured crosses.  We also show the radius to which the absolute values of the density profiles in the lower-resolution simulations agree to within 10\% of the higher-resolution result (indicated by the open squares).

Focusing first on the left column, it can be seen that in the case of the cosmologies with either a running spectral index or WDM species, that the ratios are converged with the result in the higher-resolution simulation to the smallest possible radius calculable in the lower-resolution simulation. 
If one focuses on the region between the two vertical lines, where we still trust the absolute value of the density profiles in the higher-resolution simulation, it can be seen that the ratio of the lower-resolution simulations agree extremely well with that in the higher-resolution result. 
This result is also echoed in the right column, but the ratios are more subject to shot noise due to the fewer number of haloes in this bin used to take a median. 

However, this trend breaks in the case of the SIDM cosmologies, which may not be surprising due to the fact that cores develop in the inner regions due to physical processes and not just numerical two-body relaxation effects. 
In the case of the left column and for the cosmology with the larger cross-section for interaction, it can be seen that the lower-resolution simulation fails to capture both the core-formation due to SIDM (and the resulting density excess at slightly larger radii), which is resolved in the higher-resolution simulation.  On the other hand, in the right-column the result is reversed relative to the other cosmologies. Here the absolute density profiles agree all the way to the inner-most radial bin in the case of the lower-resolution simulation, whereas the ratio is converged to the calculated convergence radius.  We speculate that the reason for this is that the energy exchange due to the particle collisions is far larger than the small changes in energy due to two-body relaxation. As such, the core formed due to these particles collisions dominates that induced due to discrete sampling effects.
The reason the ratio is not converged as well is because the core is now converged to the higher-resolution result, whereas there is an artificial core induced in the $\Lambda$CDM result.  Consequently, the suppression in the ratio underestimates the core for the lower-resolution simulation. 

These results are interesting and motivate us to examine the ratios below the convergence radius of \citet{Ludlow2019} calculated for these simulations.  For the case of SIDM cosmologies, this strategy may result in a slight underestimation of the actual (relative) effects present, as explained above.

Furthermore, these results show that when studying the effects of cosmological extensions on quantities such as the density profiles and circular velocity profiles, it may be possible to run a single high-resolution $\Lambda$CDM simulation and use the ratios from interesting cosmological extensions run at considerably lower resolution as a multiplicative adjustment to the high-resolution simulation.  This approach can potentially save a great deal of computational expense.

\subsection{Radial distribution of subhaloes}
\label{sec:rad_sub_haloes}
\begin{figure*}
    \centering
    \includegraphics[width=\textwidth]{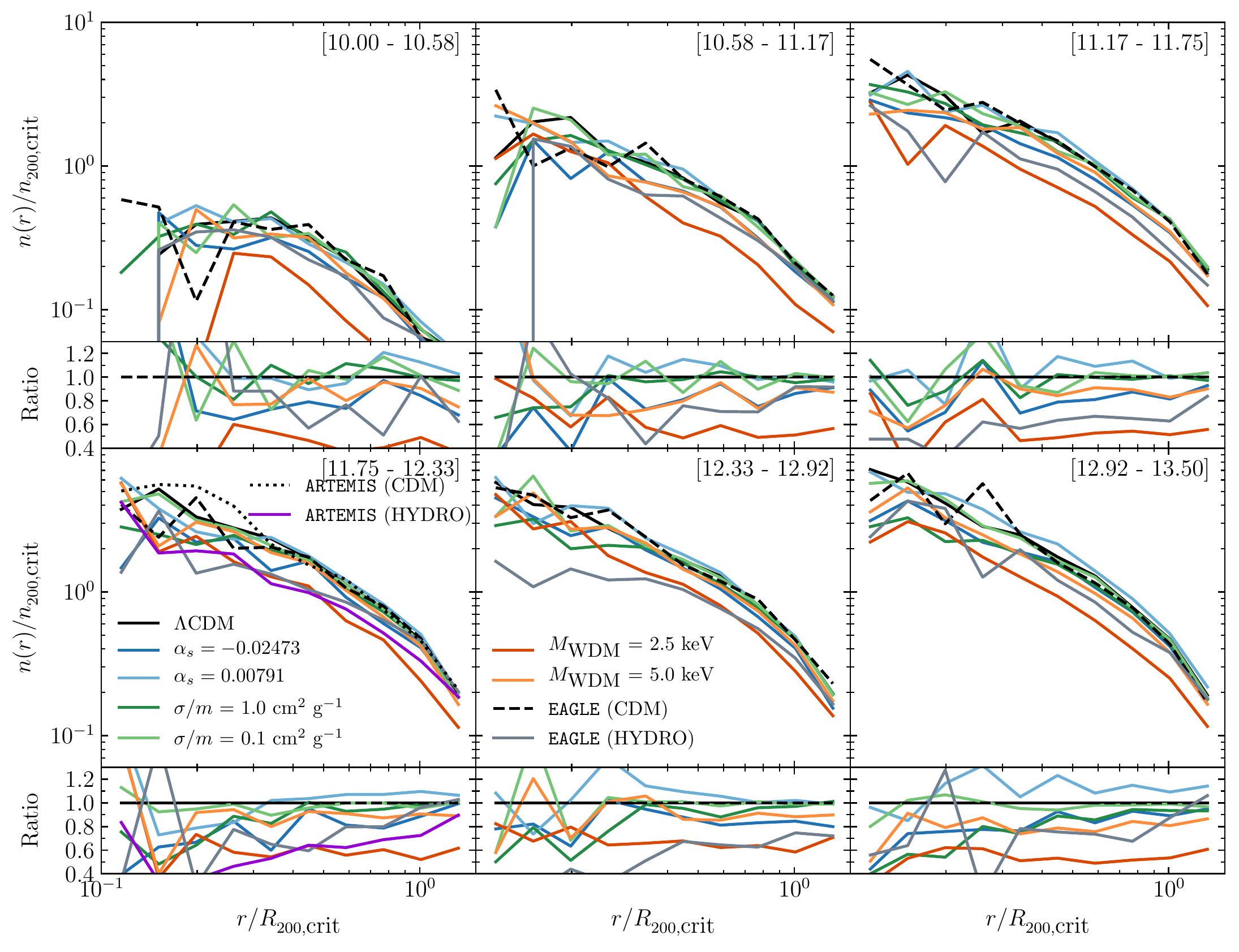}
    \vspace{-0.5cm}
    \caption{The stacked median spherically-averaged radial number density of all satellite subhaloes identified by the \texttt{SUBFIND} algorithm with a mass $> 5\:\times\:10^{8}\;\textrm{M}_{\odot}\;h^{-1}$, and $V_{\textrm{max}}$ $> 15\:\textrm{km}\:\textrm{s}^{-1}$ inside host haloes in different mass ranges (indicated in the top right-hand corner of each panel). The result in each panel is normalised with respect to the average number density of subhaloes inside $R_{200,\textrm{crit}}$ for that mass bin, in the reference cosmology. The bottom panel(s) below each plot show the result normalised with respect to the median result in the reference cosmology.}
    \label{fig:subhalo_rad_distrib}
\end{figure*}

So far we have focused on the effects that cosmological variations or baryons have on the matter distribution in host haloes.  We turn now to the effects on the subhaloes that exist within these host haloes.   In particular, we show in Fig.~\ref{fig:subhalo_rad_distrib} the median radial number density of subhaloes as a function of host halo mass. 
Note that we only include subhaloes which satisfy our $V_{\textrm{max}}$ ($> 15\;\textrm{km}\;\textrm{s}^{-1}$) and $M_{\textrm{sub}}$ ($> 5\:\times\:10^{8}\;\textrm{M}_{\odot}\;h^{-1}$) convergence criteria.
We normalise the resultant value in each mass bin, for each cosmology, by the average number density of subhaloes inside $R_{200,\textrm{crit}}$ calculated for the reference cosmology. 
The bottom panels show the results when normalised to the reference $\Lambda$CDM cosmology in that mass bin.

It can be seen that the largest effect in all mass bins comes from the cosmology with the lightest WDM mass.  
This makes sense, as this model has the largest suppression of the SMF (particularly at the low-mass end, which will dominate the signal seen here due to the steepness of the SMF).
Adopting a 2.5 (5 keV) WDM particle mass, we see a $40\%$ ($20\%$) suppression in the number density of subhaloes, which is roughly independent of distance from the centre of the host halo (this is in agreement with, for example, the suppression seen in the number of satellites in \citealt{Lovell2017b}).

A similar effect is seen in the cosmology which has a negative running, in that, for all mass bins there is a suppression at all radii compared with the result in the standard cosmology.
The magnitude of the suppression in the number density ($\approx 20\%$ in this case) is similar to that seen for the 5 keV WDM model.
Conversely, for the positive-running cosmology, there does not appear to be much of an effect relative to the reference cosmology, with there being only a very mild hint of an enhancement in the number of subhaloes at all radii in the larger mass bins.

Examining the result for the SIDM cosmologies, in the $\sigma/m = 0.1\:\textrm{cm}^2\:\textrm{g}^{-1}$ cross-section there appears to be no discernible effect in any of the mass bins.  When examining the more extreme SIDM model, there is also little to no effect in the low mass bins.  However, at the high-mass end it can be seen that there is a clear suppression in the number density of subhaloes in the inner regions of these haloes, which increases with increasing host halo mass and with decreasing distance to the centre of the host halo.  
As discussed in Section \ref{sec:global_properties}, this effect is the result of heat transfer (the dark matter equivalent of thermal evaporation) due to scattering between `hot' host dark matter particles and `cooler' subhalo/satellite particles.
This effect happens in conjunction with enhanced tidal stripping, due to the cored density profiles of these subhaloes \citep{Penarrubia2010}. 
This effect was previously observed in \cite{Vogelsberger2012} for a MW-type halo (see also \citealt{Nadler2020a, Banerjee2020}).

These results are interesting as they show that, in the case of the WDM cosmologies, there is an almost systematic suppression in the number density of subhaloes as a function of radius in all mass bins.
This means that, in principle, a technique where one simply counts the total number of subhaloes (satellite galaxies) could be used to rule out the most extreme WDM models, which as discussed in Section \ref{sec:sims}, is a technique already being used to place constraints on WDM masses \citep[e.g.][]{Lovell2014}. 
However, this is slightly more complicated in the case of the less extreme WDM model, as this plot shows that the effects present in a cosmology with a negative running of the spectral index are highly degenerate to those in this WDM cosmology\footnote{Note that even the more extreme WDM model could effectively masquerade as a less extreme one if one also invokes a positive running scalar spectral index.}.
As such, it would be potentially difficult to disentangle these two effects through this type of observation alone.
However, going back to the density and circular velocity profiles of the haloes which host these satellites, there are more apparent differences in the underlying matter distribution which could then be used to differentiate these two cosmologies. 
Similarly, in the case of the most extreme SIDM cosmology, counting subhaloes could also potentially be used to constrain these models, in that for high-mass haloes, there appears to be a clear radial dependence in the suppression of the number density of subhaloes.
This is a result which is unique among the cosmological extensions examined here.
It is worth noting, however, that these are the signals predicted in the absence of baryonic physics.  Including baryonic effects may modify many of these signatures (e.g. \citealt{Richings2020}), making potential constraints with these observations more difficult. 

We can get a sense of the potential impact of baryons by examining the hydrodynamical simulations.  For \texttt{EAGLE}, there is a suppression in the number density of subhaloes at virtually all radii and across all mass bins.  
We speculate that this suppression is due not only to change in subhalo mass and maximum circular velocity, but is also a result of the central region of the host halo being considerably more dense (due to stars and associated contraction of the dark matter), which will result in enhanced tidal stripping of the satellites in the hydrodynamical simulations \citep[see e.g.][]{Samuel2020}.
The baryonic effects are therefore similar to those seen for almost all the cosmological extensions explored here (with the exception of the positive-running cosmology).
This is also true for the \texttt{ARTEMIS} simulations (lower-left panel of the plot), which shows an almost systematic suppression in the number density of subhaloes. 
It does appear however that there is a slight radial dependence to this suppression (which is also hinted at in the \texttt{EAGLE} result). 

\subsection{Concentration--mass relation}
\label{sec:conc_mass}
\begin{figure}
    \centering
    \includegraphics[width=\columnwidth]{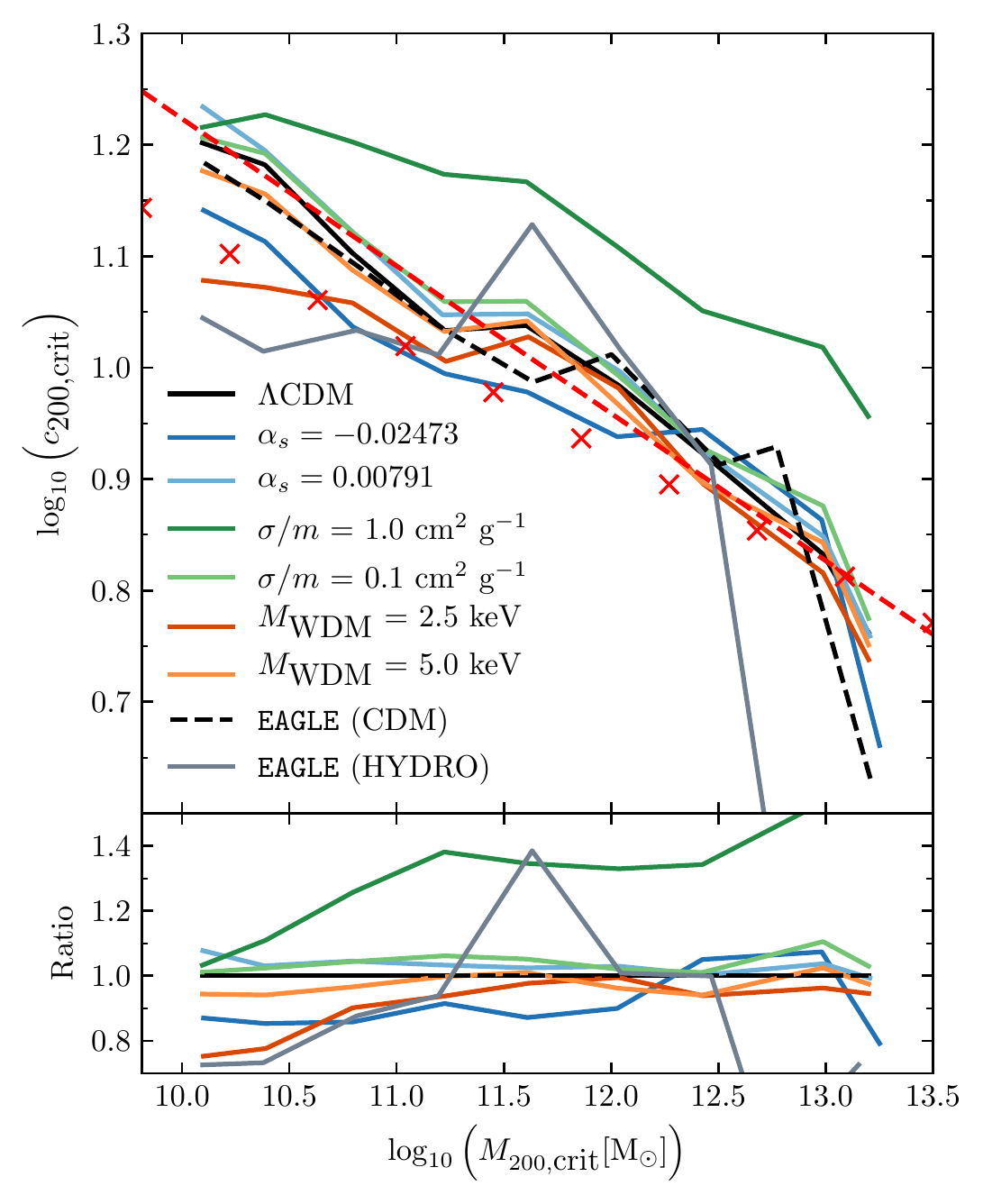}
    \vspace{-0.5cm}
    \caption{Top: the concentration-mass relation computed for stacks of host haloes in the mass range $9.81 \leq M_{200,\textrm{crit}}/\textrm{M}_{\odot} \leq 13.0$ for each cosmology at $z = 0$. 
    The haloes are stacked into 9 equally spaced logarithmic mass bins. 
    The concentration parameter is calculated for each stack by computing the logarithmic slope of the stacked density profile, and finding the radius for which this value is equal to -2. Bottom: the $c$--$M$ relation normalised with respect to the result in the reference simulation.  
    The red crosses corresponds to the empirical relation found in \citet{Dutton2014}; with the dashed red line corresponding to this relation, with the parameters tuned to best-fit the reference simulation in this suite.
    }
    \label{fig:conc_mass}
\end{figure}

The concentration--mass relation is of particular interest as it has been shown through cosmological simulations that the inner density profiles of haloes are representative of the conditions of the Universe at the formation time of the halo, at least in a $\Lambda$CDM model. This result (illustrated by \citealt{Navarro1996} for example, but confirmed by many additional studies), implies that low-mass haloes, which form first, have a higher central concentration than higher-mass haloes, which tend to assemble most of their mass late on.
This is just due to the fact that low-mass haloes formed at a redshift when the mean density of the universe was higher.

It has also been shown that the concentration of low-mass haloes is affected in a WDM cosmology \citep[e.g.][]{Eke2001, Bode2001, AvilaReese2001, Schneider2015, Ludlow2016}, with lower-mass haloes having lower-concentrations compared with the corresponding halo mass CDM haloes, with this difference decreasing with increasing halo mass.
Alongside this, it has been shown that a running scalar spectral index can also alter the $c$--$M$ relation of high-mass haloes \citep{Fedeli2010, Stafford2020}.
It has been shown by \cite{Despali2019} that allowing for self-interactions between the dark matter, also tends to slightly alter the $c$--$M$ relation.  As such, it is interesting to compare and contrast all of these different models in a consistent manner to see the relative effects they have on this important statistic. 

It is commonplace when computing the concentration of a halo to fit a Navarro, Frenk \& White (NFW) profile \citep{Navarro1996} to the halo (or equivalently an Einasto profile \citealt{Einasto65}).  The NFW profile has two free parameters: a scale radius and a scale density, or equivalently to this, a halo mass and a concentration.  The concentration is defined as $c_{\Delta} \equiv R_{\Delta}/r_s$ ($r_s \rightarrow r_{-2}$ in the case of the Einasto profile), where $\Delta$ corresponds to a choice of spherical overdensity (which in this study is set equal to 200, with respect to the critical density) and $r_s$ is the scale radius, defined as the radius where the logarithmic slope of the density profile is $-2$.

In this study we take a different approach, as we want to consistently compare the concentration parameter obtained as a function of mass for the different cosmologies examined here. This is not possible using either the Einasto or NFW forms for the SIDM models, as the SIDM models produce prominent cores in the dark matter density profiles (see for example Fig.~\ref{fig:density_profiles_box}). 
One could instead use a Burkert profile \citep{Burkert1995}, which is able to fit the cored-region in these dark matter haloes. 
However, this comes at the expense of poorly fitting the outer regions of SIDM haloes (see e.g. \citealt{Rocha2013}) and would also provide a poor fit to the haloes in the other cosmologies.
Therefore, we instead take a non-parametric approach and numerically evaluate the logarithmic slope of the computed density profiles using a 2nd-order \cite{Savitzky1964} filter, smoothing over the 7 nearest bins.  We then identify the radius for which the logarithmic slope equals $-2$. Thus, our non-parametrically estimated concentration parameter is defined in a way which is consistent with the standard definition of concentration using either the NFW or Einasto parametric forms, but it means the same technique can be applied to all the simulations. 
To check our method, we have compared the concentration parameter obtained for stacked density profiles in 6 different mass bins using this method to that obtained by fitting an NFW and Einasto profile to the stacks of haloes in the reference $\Lambda$CDM model.  Our method provides consistent concentrations with those estimated using the NFW or Einasto forms.

We stack haloes with masses between $9.81 \leq M_{200,\textrm{crit}}/\textrm{M}_{\odot} \leq 13.0$ in 9 equally spaced logarithmic mass bins, with the lower mass limit being chosen in order to only include haloes with a minimum of 5000 particles inside $R_{200, \textrm{crit}}$.
The spherically-averaged dark matter-only density profiles are computed in 32 logarithmically spaced radial bins, spanning a radial range: $-2.5 \leq \log_{10}(r/R_{200,\textrm{crit}}) \leq 0$, as in \cite{Ludlow2016}. 

The resultant $c$--$M$ relation can be seen in Fig.~\ref{fig:conc_mass}, where the bottom panels in this plot is the $c$--$M$ relation for each cosmology normalised to the result in the reference $\Lambda$CDM simulation.
This figure shows that qualitatively lower-mass haloes are more centrally concentrated, with the concentration steadily declining with increasing halo mass, irrespective of the details of cosmology or baryons.
However, there are clear quantitative differences between the individual models. 
For example, as expected, the two different WDM models predict the low-mass haloes to be less centrally-concentrated compared to the standard model result, with this effect being more extreme for the lighter 2.5 keV model.
This cosmology predicts a $\approx20\%$ suppression in the concentration levels of $10^{10}$ M$_{\odot}$ haloes. 
This is a similar level of suppression to that found in previous studies such as \citet{Schneider2015} and \citet{Ludlow2016}, who adopted similar WDM masses ([3.0, 3.3] keV respectively).

The magnitude of the effect on the $c$--$M$ relation in a cosmology with a running spectral index scales with the magnitude of the running parameter.
For example, the cosmology with a negative running produces a suppression of around 10\% in the central concentrations of haloes of fixed mass, with this effect extending over almost the entire mass range covered in this study. 
Whereas a positive running cosmology does not appear to have much of an effect in this mass range, confined to the few percent level.
These results extend the trends seen previously in \cite{Stafford2020} to lower halo masses with, for example, lower mass haloes being less centrally concentrated in a cosmology with a negative running spectral index relative to the standard model.

The largest effect seen is in the most extreme SIDM model, which produces a large {\it increase} in the concentration at fixed mass.  There also appears to be a change in slope for the $c$--$M$ relation in the more extreme SIDM cosmology, with the offset between the standard model result and the SIDM result increasing with increasing mass.  Naively, this appears to be inconsistent with the cores seen in the density profiles shown earlier (Fig.~\ref{fig:density_profiles_box}).  The increase in the concentration, or more specifically the decrease in the radius where the logarithmic slope equals $-2$, is a consequence of the enhancement in the density just outside the core.  The enhancement causes the transition radius from a constant density core in the inner regions of a SIDM halo to an NFW-like logarithmic slope of -3 in the outer regions to move inwards.

When examining the result in the \texttt{EAGLE} Recal simulation, there appears to be a change in slope for the $c$--$M$ relation, compared to the result in the dark matter-only simulation. 
In particular the $c$--$M$ relation in the full hydrodynamic simulation is somewhat flatter, in agreement with that found in \cite{Schaller2015}. This is because the decrease in $M_{200, \textrm{crit}}$, seen in Fig. \ref{fig:frac_mass_diff_haloes}, also results in a decrease in a haloes' $R_{200, \textrm{crit}}$. This causes a decrease in concentration, preferentially at low halo masses. 

Previous studies have shown that simple power-laws or functions are able to describe well the average $c$--$M$ relation of haloes over a narrow range in halo masses \citep[e.g.][]{Avila-Reese1999, Neto2007, Duffy2008, Dutton2014, Child2018}.
We highlight one such result here, that by \cite{Dutton2014}, who found that the $c$--$M$ relation of haloes in a \cite{Ade2014} cosmology is well described by a power-law of the form:
\begin{ceqn}
\begin{align}
\label{eq:dutton_conc}
\log_{10}(c) = a + b\log_{10}\left(\frac{M}{M_\textrm{pivot} \;[\textrm{M}_{\odot}\:h^{-1}]}\right),
\end{align}
\end{ceqn}
 
 where $a$, $b$ are fitting constants, found to be 0.905, -0.101 respectively in \cite{Dutton2014}, $M_\textrm{pivot}$ is a reference mass-scale, set to be 10$^{12}$ $\textrm{M}_{\odot}$ $h^{-1}$. We show this result as red crosses in Fig. \ref{fig:conc_mass}. We also show an updated fit to our reference cosmology, which provides best-fitting parameters $a$ = 0.936, $b$ = -0.132. 
 The reasons for the slight change in these parameters are two-fold. Firstly, the cosmologies are slightly different, for example the values for $\Omega_\textrm{m}$ and $\sigma_{8}$ which the $c$--$M$ relation is particularly sensitive to \citep{Duffy2008, Dutton2014}.
 Also, our method of taking the logarithmic slope of the median density profile in a mass bin predicts slightly higher values for the concentration than what one typically gets when fitting an NFW profile to that same stack. This is particularly true at low-masses, which is where the two best-fitting relations differ most.
 
\FloatBarrier
\section{Discussion and Conclusions}
\label{sec:conclusions}

In this study we have systematically explored different proposed mechanisms for altering the distribution of matter on small scales.  This includes alterations to the standard model of cosmology or including baryonic effects within the standard model (but not yet both simultaneously).  These mechanisms are motivated in part by apparent tensions between the standard model and observational data in the local Universe (see the Introduction).  

In terms of the alterations to the standard model, we consider three scenarios: warm dark matter (WDM), self-interacting dark matter (SIDM), and a running of the scalar spectral index of the power spectrum of primordial density fluctuations.  These cosmological alterations come with additional free parameters (e.g., WDM particle mass, SIDM cross-section, value of the running) and we use current observational constraints to guide our choice of the parameter values.  To characterise the potential role of baryons within the standard model, we use the high-resolution \texttt{EAGLE} `Recal' run and the \texttt{ARTEMIS} suite of zoom-in simulations of Milky Way-mass haloes.  Using all of the simulations, we compare and contrast the effects of altering the standard model in different ways and we frame this within the context of the potential role of baryons, highlighting potential degeneracies between the different effects for different observables.

The main findings of our study may be summarised as follows:
\begin{itemize}
    \item Baryon physics, WDM, and a running spectral index alter the spherical-overdensity (SO) halo mass function (Fig. \ref{fig:HMF}) in similar ways and are expected to be degenerate if they are constrained solely using this metric. 
    A cosmology where the DM is allowed to self-interact (SIDM) is the only cosmological extension examined here which has no significant effect on the SO halo mass function. 
    There are, however, noticeable effects present when examining the subhalo population in SIDM cosmologies (Fig. \ref{fig:sub_mass_func}).   Specifically, while the central subhalo mass function resembles the SO mass function strongly, the satellite subhalo mass function can be strongly suppressed, particularly for satellites that are located close to the centres of their host halo (Fig. \ref{fig:subhalo_rad_distrib}).  This effect is plausibly due both to a form of evaporation in which energetic DM particles from the host halo scatter off satellite DM particles (and thus unbinding them) and enhanced tidal stripping as a result of the cored density profiles of SIDM subhaloes.
 
    \item When examining the radial dark matter distribution within host haloes, WDM and a (negative) running spectral index can behave similarly (Figs. \ref{fig:density_profiles_box}, \ref{fig:circ_vel_profile}).
    This degeneracy is only prevalent for haloes with mass $\lesssim 10^{11.75}\;\textrm{M}_{\odot}$, however, with there being no significant suppression in the density profiles in the WDM cosmologies for haloes with mass greater than this, but still an $\approx$ 10\% suppression in the inner regions for the negative running cosmology.
    The SIDM cosmologies produce large cores, strongly distinguishing them from any of the other cosmological extensions.
    These results are echoed in the concentration--mass relation of haloes (Fig. \ref{fig:conc_mass}).
    Consistent with previous works, there is a large enhancement in the dark matter density profile for the hydrodynamical simulations, shown both for the \texttt{EAGLE} Recal simulation, as well as the \texttt{ARTEMIS} zoom-in simulations.
    Interestingly, among the cosmological extensions we also find that a positive running scalar spectral index is able to produce an enhancement in the dark matter density profiles relative to the $\Lambda$CDM case, but at a much reduced level (around 10\%) for haloes of mass $ \leq 10^{12.33}\;\textrm{M}_{\odot}$.
    
    \item We have shown that for cosmological extensions such as a running scalar spectral index and WDM, the \textit{relative} effects on the density profiles with respect to the standard model are reliably followed down to significantly smaller radii than standard `convergence' radii (for absolute quantities) would suggest (Fig. \ref{fig:conv_test_ratios}).  This potentially allows one to save a great deal of computational expense, by running only a single $\Lambda$CDM simulation at very high resolution and then using the ratios of cosmological extensions to $\Lambda$CDM from lower-resolution simulations to produce high-resolution non-standard cosmologies.

    \item There is a strong degeneracy for the suppression in the radial distribution of subhaloes in their host haloes for WDM, running, and hydrodynamical effects (Fig.~\ref{fig:subhalo_rad_distrib}).
    For example, the results for the 5.0 keV WDM model are nearly identical to those with the negative running cosmology.
    In the high halo-mass bins ($M \gtrsim 10^{11.75}\;\textrm{M}_{\odot}$) there is also a suppression in the number density of subhaloes in the SIDM cosmology with the larger cross-section for interaction, due to evaporation of the satellites and enhanced tidal stripping of cored subhaloes. 

\end{itemize}

The different cosmological extensions explored here are able to produce large, and potentially measureable effects on the structure which forms in the Universe.  
All of the cosmological extensions examined are within current observational constraints, and are able to potentially help alleviate some of the small-scale crises which (may) exist with the standard model of cosmology. 
An interesting problem arises, however, in that some of the models have highly degenerate effects.

As such, when making constraints on these different cosmological extensions, it is potentially necessary to combine probes which examine different aspects of haloes and their satellite subhaloes in order to break degeneracies between the different extensions.  This is because, in detail, the different extensions do have different mass and radial dependencies which can be exploited to simultaneously constrain them.

However, it is important to note that this study has focused on dark matter-only simulations when extending $\Lambda$CDM. 
We examined the \texttt{EAGLE} `Recal' and \texttt{ARTEMIS} simulations to compare the results from these gravity-only simulations to those from full hydrodynamical simulations (in the context of the standard model).  We find that baryonic physics can also be highly degenerate with varying the cosmological model.
Furthermore, there are inherent differences in the results predicted by the inclusion of baryonic physics, due to the uncertainty associated with the subgrid implementation of this physics. This is highlighted by the differences present between the \texttt{ARTEMIS} results and the \texttt{EAGLE} results.  Even though these simulations have the same subgrid physics implementations, they were calibrated slightly differently, leading to somewhat different results.  While it is clear that baryonic effects are important, how robust the predictions are remains an open question.  Since feedback efficiencies cannot be predicted from first principles and must therefore be calibrated against some set of observable properties of galaxies \citep{Schaye2015}, there is a danger that we may come to erroneous conclusions about the role of baryons if those calibration observables also depend on cosmology.  

Thus, given the degeneracies that exist between cosmological extensions and also that of baryon effects and the uncertainties inherent to galaxy formation modelling, it seems the best approach is to \textit{simultaneously} explore variations to cosmology and baryon physics, using a multi-observational probe approach to constrain both.  We are adopting such a strategy in forthcoming work.

\section*{Acknowledgements}
The authors thank the referee for helpful comments which improved the paper.
The authors thank Joop Schaye for helpful comments.  SGS thanks Matthieu Schaller for assistance with the EAGLE data.
SGS acknowledges an STFC doctoral studentship. This project has received funding from the European Research Council (ERC) under the European Union's Horizon 2020 research and innovation programme (grant agreement No 769130). AR is supported  by the European Research Council's Horizon 2020 project `EWC' (award AMD-776247-6). This work used the DiRAC@Durham facility managed by the Institute for Computational Cosmology on behalf of the STFC DiRAC HPC Facility. The equipment was funded by BEIS capital funding via STFC capital grants ST/P002293/1, ST/R002371/1 and ST/S002502/1, Durham University and STFC operations grant ST/R000832/1. DiRAC is part of the National e-Infrastructure.

\section*{Data Availability Statement}
The data underlying this article will be shared on reasonable request to the corresponding author.
%%%%%%%%%%%%%%%%%%%%%%%%%%%%%%%%%%%%%%%%%%%%%%%%%%



\bibliographystyle{mnras}
\bibliography{Bibliography.bib} % if your bibtex file is called example.bib

%%%%%%%%%%%%%%%%%%%%%%%%%%%%%%%%%%%%%%%%%%%%%%%%%%

%%%%%%%%%%%%%%%%% APPENDICES %%%%%%%%%%%%%%%%%%%%%

\appendix
\FloatBarrier
\section{Convergence tests}
\label{sec:convergence_tests}
Here we determine the mass and maximum circular velocity limits for the subhaloes that we include in the statistics examined in this paper.

\begin{figure}
    \centering
    \includegraphics[width=\columnwidth]{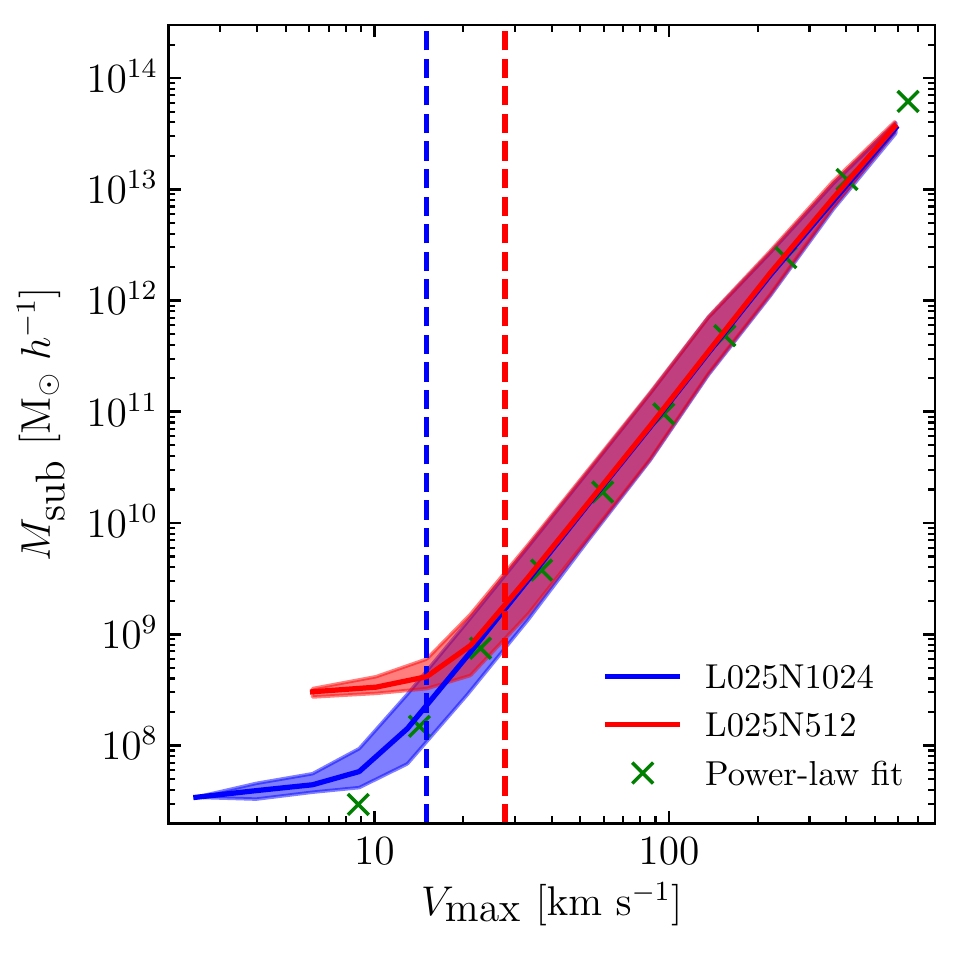}
    \vspace{-0.5cm}
    \caption{$M_{\textrm{sub}}$-$V_{\textrm{max}}$ relation for subhaloes in simulations ran at two different resolution levels indicated by colour. These simulations correspond to the reference cosmology used through this study. The green crosses correspond to the power-law fit calculated using equation \ref{eq:boylan-kolchin_fit}, as done in \citet{Boylan-Kolchin2010}. The vertical dashed lines correspond to the resolution limit for these simulations, beyond which numerical effects become important. The shaded regions show the 16th and 84th percentiles.}
    \label{fig:conv_test_vmax-msub}
\end{figure}

\begin{figure}
    \centering
    \includegraphics[width=\columnwidth]{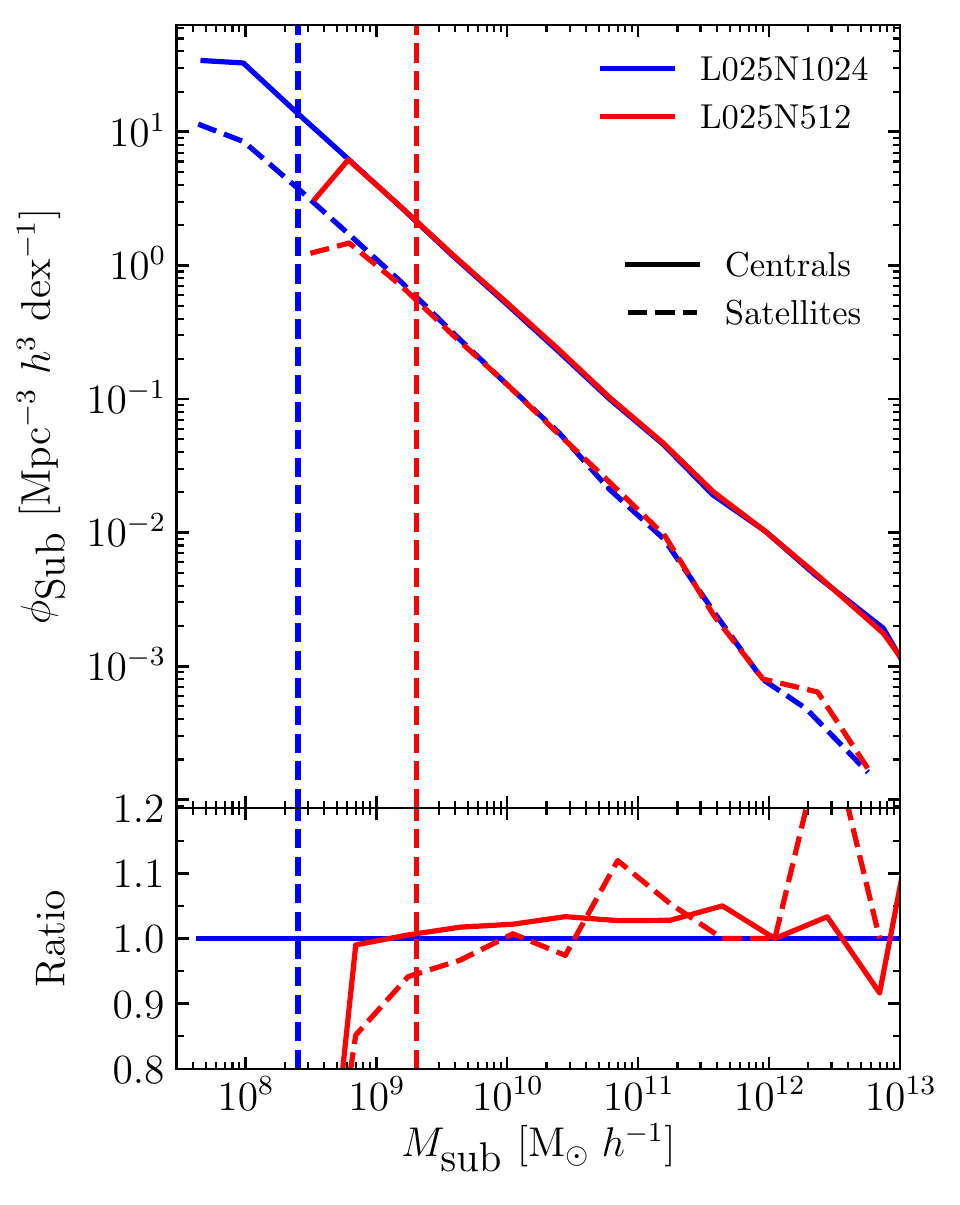}
    \vspace{-0.5cm}
    \caption{Top: the subhalo mass function for two different resolution tiers (indicated by colour). The solid line corresponds to this quantity calculated for centrals, with the dashed line corresponding to satellites. Bottom: the above result normalised with respect to the result in the higher resolution simulation. The vertical dashed line corresponds to the lower-mass limit for each simulation, calculated using the mass at which the lower-resolution simulation diverges from the higher-resolution case.}
    \label{fig:conv_test_msub_func}
\end{figure}

We examine the $M_{\textrm{sub}}$-$V_{\textrm{max}}$ relation, shown in Fig.~\ref{fig:conv_test_vmax-msub}, for two different resolution levels. Following \cite{Boylan-Kolchin2010}, we fit a power-law to this relation of the form:
\begin{ceqn}
\begin{align}
\label{eq:boylan-kolchin_fit}
  \log_{10}\left(M_{\textrm{sub}} \; [\textrm{M}_{\odot}\:h^{-1}]\right) = 11.04 + 3.38\log_{10}\left(\frac{V_{\textrm{max}}}{100 \; \textrm{km s$^{-1}$}}\right),  
\end{align}
\end{ceqn}
which is shown as crosses in the plot. 
It can be seen that the higher-resolution simulations used throughout this study follow the power-law down to a $V_{\textrm{max}} \approx$ 15 km s$^{-1}$ ($\approx 3 \times 10^{8}\;\textrm{M}_{\odot}\;h^{-1}$), before the turn-up indicative of resolution effects. 
The same effect is seen in the lower-resolution simulation, but at a lower value for $V_{\textrm{max}} \approx 28$ km s$^{-1}$. 
In addition, we plot the subhalo mass function, shown in Fig.~\ref{fig:conv_test_msub_func}, for both centrals and satellites, for two tiers of resolution. 
Here, we examine at what particle limit does the subhalo mass function in the lower-resolution simulation start to significantly differ from the higher-resolution simulation.
This is illustrated in the bottom panel of this figure, which shows the ratio of the lower-resolution simulation, with respect to the higher-resolution simulation. 
We place a conservative lower-limit on the number of particles, for which we deem this quantity to be converged, which is indicated by the vertical dashed lines, and corresponds to subhaloes containing at least 200 particles.
This corresponds to a mass-limit of $2.5 \times 10^8$ M$_{\odot}\;h^{-1}$.
Therefore, throughout this paper, we only select subhaloes with $M_{\textrm{sub}} > 5\times10^{8}$ M$_{\odot}\;h^{-1}$, and $V_{\textrm{max}} > 15$ km s$^{-1}$, satisfying both convergence tests.

\begin{figure}
    \centering
    \includegraphics[width=\columnwidth]{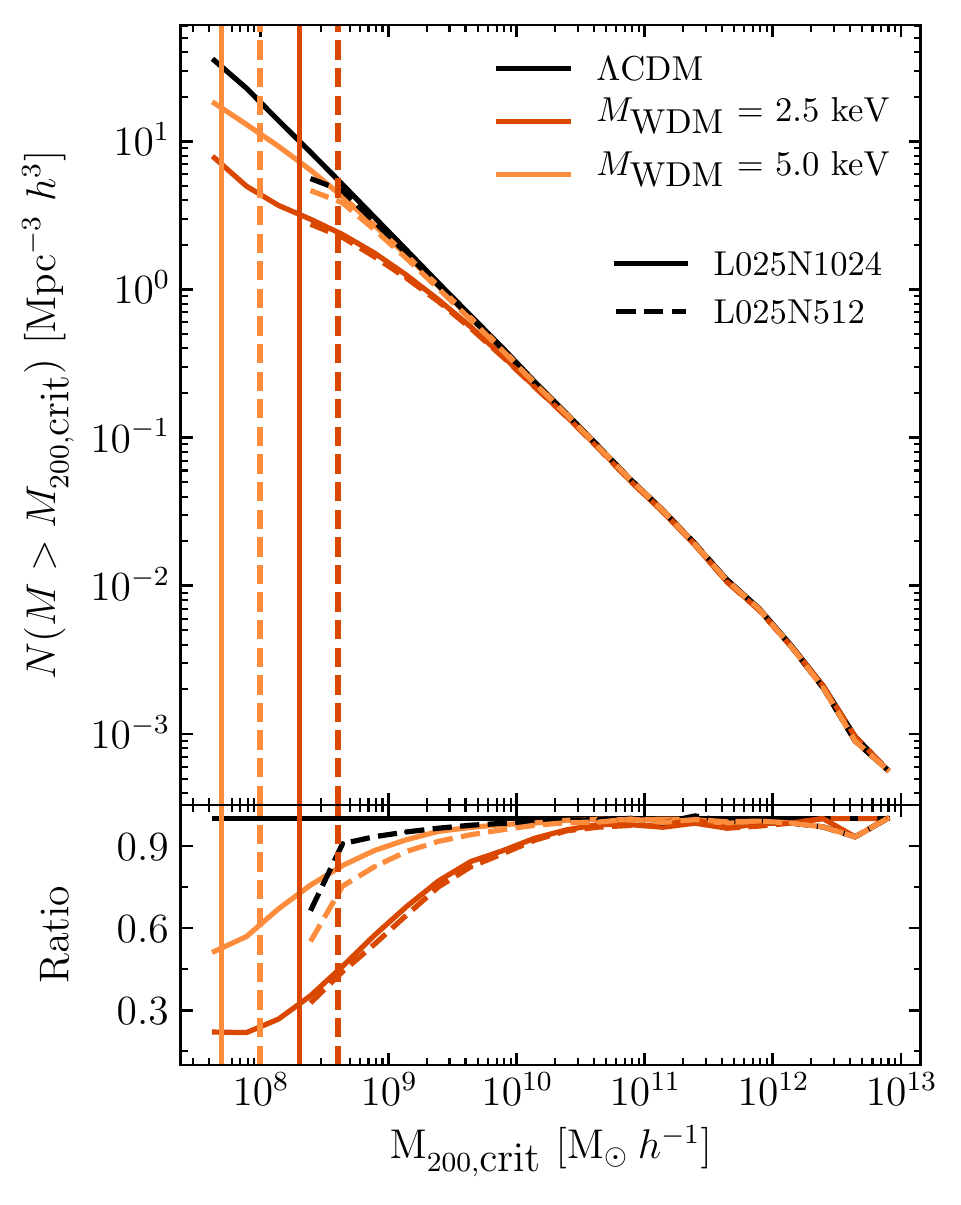}
    \vspace{-0.5cm}
    \caption{Top: cumulative abundance of subhaloes above a mass $M$ for the different WDM cosmologies, along with the reference cosmology. Bottom: the above result normalised with respect to the reference simulation. The vertical dashed lines indicate the mass cut advocated by \citet{Wang2007} for the two different WDM models.}
    \label{fig:spurious_halo_check}
\end{figure}

As mentioned in the main text, it was shown by \cite{Wang2007} that spurious fragmentation of filaments is a significant problem in WDM simulations due to discreteness effects.  They derived an empirical mass cut to remove these spurious objects: $M_{\textrm{lim}}\:=\:10.1\overline{\rho}d k_{\textrm{peak}}^{-2}$.
Where $\overline{\rho}$ is equal to the mean density of the universe ($\Omega_{\textrm{m}} \rho_{\textrm{crit}}$), $d$ is the mean inter-particle separation, and $k_{\textrm{peak}}$ is equal to the maximum of the dimensionless linear theory matter power spectrum ($\Delta^2(k) = k^3P(k)/(2\pi^2)$). 
Therefore, we show in Fig.~\ref{fig:spurious_halo_check} the cumulative abundance of haloes above a certain mass $M$, with this mass limit shown with a vertical dashed line.
We plot this for both of the WDM models simulated as part of this study, at two different resolutions, illustrating the resolution dependence of this effect.
The resultant mass-cut advised is equal to $\approx$ (0.5, 2) $\times 10^{8}$ M$_{\odot}\:h^{-1}$ for the the $M_{\textrm{WDM}}$ = 5.0, 2.5 keV cosmologies respectively. 
Both of these mass limits lie below the resolution cut we derived above.  We therefore do not need to make an additional cut for the WDM simulations.

One additional note worth mentioning here is the apparent lack of spurious haloes in the lower-resolution simulation, which at first sight seems to contradict the expectation that they should be more prevalent at lower resolution. 
However, the reason for this behaviour is that the WDM models which we explore have a cut-off in their initial matter power spectrum that results in a sufficiently large $k_{\rm{peak}}$ that the spurious halo mass limit is barely resolved in the lower resolution simulation. For example, $M_{\rm{lim}}$ corresponds to $\approx$ 10 (40) particles for the $M_{\rm{WDM}}$ = 5.0 keV (2.5 keV) model. As such, the vast majority of these haloes will not be identified as FoF groups (we require there to be at least 20 particles in identified FoF groups), which is why the characteristic turn up in the mass function is not seen for the lower-resolution simulations.

%%%%%%%%%%%%%%%%%%%%%%%%%%%%%%%%%%%%%%%%%%%%%%%%%%

% Don't change these lines
\bsp	% typesetting comment
\label{lastpage}
\end{document}